\documentclass[reprint, unsortedaddress, nobibnotes, amsmath,amssymb, aps, pre, superscriptaddress, floatfix, longbibliography]{revtex4-2}

\usepackage{graphicx}
\usepackage{bm}
\usepackage{xfrac}
\usepackage{mathptmx}
\usepackage{soul}
\usepackage{xcolor}
\usepackage{multirow}
\usepackage{makecell}
\usepackage{hyperref}
\hypersetup{
    colorlinks,
    linkcolor={blue!40!black},
    citecolor={blue!100!black},
    urlcolor={purple!100!black}
}
\usepackage[table]{colortbl}

\usepackage{anyfontsize}

\newcommand{\Aave}{\ensuremath{\ave{A}}}

\newcommand{\Secref}[1]{Sec.~\ref{#1}}
\newcommand{\Secsref}[1]{Secs.~\ref{#1}}
\newcommand{\Appref}[1]{Appendix~\ref{#1}}
\newcommand{\Appsref}[1]{Appendices~\ref{#1}}

\newcommand{\wt}{\widetilde}
\newcommand{\ave}[1]{{\left<#1\right>}}
\newcommand{\save}[1]{{\langle{#1}\rangle}}
\newcommand{\inv}[1]{\frac{1}{#1}}
\DeclareSymbolFont{newfont}{OML}{cmm}{m}{it} 
\DeclareMathSymbol{\Varrho}{3}{newfont}{37}
\newcommand{\abs}[1]{{\left|#1\right|}}
\newcommand{\rmd}{\text{d}}

\newcommand{\tarriv}{{\ensuremath{u}}}
\newcommand{\twait}{{\ensuremath{w}}}
\newcommand{\taudim}{\ensuremath{s}}
\newcommand{\smin}{\taudim_\downarrow} 
\newcommand{\smax}{\taudim_\uparrow}
\newcommand{\taumin}{\tau_\downarrow}
\newcommand{\taumax}{\tau_\uparrow}
\newcommand{\veldim}{\ensuremath{v}}
\newcommand{\vmin}{\veldim_\downarrow}
\newcommand{\vmax}{\veldim_\uparrow}
\newcommand{\omegadim}{\ensuremath{o}}
\newcommand{\Omegadim}{\ensuremath{S}}
\newcommand{\unifwidth}{\sigma} 

\newcommand{\constfactor}{\ensuremath{\kappa}}
\newcommand{\rate}{\ensuremath{\upsilon}}
\newcommand{\exprate}{\ensuremath{\lambda}}

\newcommand{\Phiave}{\ensuremath{\ave{\Phi}}}
\newcommand{\Phirms}{\ensuremath{\Phi}_\text{rms}}
\newcommand{\Phiwt}{\ensuremath{\widetilde{\Phi}}}

\newcommand{\Eqref}[1]{Eq.~\eqref{#1}}
\newcommand{\Eqsref}[1]{Eqs.~\eqref{#1}}
\newcommand{\Figref}[1]{Fig.~\ref{#1}}
\newcommand{\Figsref}[1]{Figs.~\ref{#1}}

\newcommand{\ccNoPdf}{\cellcolor{gray!50}}
\newcommand{\ccNosAve}{\cellcolor{gray!35}}
\newcommand{\ccNoScal}{\cellcolor{white!15}}
\newcommand{\cclnCorr}{\cellcolor{orange!15}}
\newcommand{\ccYesScal}{\cellcolor{orange!25}}

\begin{document}

\preprint{APS/123-QED}

\title{Long-range correlations with finite-size effects from a superposition of uncorrelated pulses with power-law distributed durations} %

\author{M.~A.~Korzeniowska}
 \email{magdalena.a.korzeniowska@uit.no}
\author{O.~E.~Garcia}
 \email{odd.erik.garcia@uit.no}
\affiliation{Department of Physics and Technology, UiT The Arctic University of Norway, N-9037 Troms{\o}, Norway.}

\date{\today}

\begin{abstract}
Long-range correlations manifested as power spectral density scaling $1/f^\beta$ for frequency $f$ and a range of exponents $\beta$ are investigated for a superposition of uncorrelated pulses with distributed durations $\tau$. 
Closed-form expressions for the frequency power spectral density are derived for a one-sided exponential pulse function and several variants of bounded and unbounded power-law distributions of pulse durations ${P_\tau(\tau)\sim1/\tau^\alpha}$ with abrupt and smooth cutoffs.
The asymptotic scaling relation $\beta=3-\alpha$ is demonstrated for $1<\alpha<3$ in the limit of an infinitely broad distribution $P_\tau(\tau)$.
Logarithmic corrections to the frequency scaling are exposed at the boundaries of the long-range dependence regime, $\beta=0$ and $\beta=2$.
Analytically demonstrated finite-size effects associated with distribution truncations are shown to reduce the frequency ranges of scale invariance by several decades. The regimes of validity of the $\beta=3-\alpha$ relation are clarified.
\end{abstract}

\maketitle

\section{Introduction}\label{sec:intro}

Coherent propagating structures, such as solar phenomena and turbulent flows, are robust examples of naturally occurring scale invariance, long-range correlations and $1/f$ noise
\cite{1999-Boffetta-PhysRevLett.83.4662, 1999-Chang-ApSS264303C, 1999-Veltri-PPCF41A787V, 2006-Arcangelis-PhysRevLett.96.051102, 2007-Uritsky-PhysRevLett.99.025001, 2007-Kiyani-PhysRevLett.98.211101, 2019-Aschwanden, 2021-Aschwanden-Wit, 2019-Pereira-PhysRevE.99.023106}.
Phenomenological modelling of the self-similarity characteristics of fluctuating systems has canonically been done using stochastic processes describing a superposition of uncorrelated pulses \cite{2000-pecseli-book, 2005-Lowen-book, 2011-Aschwanden-book, 2016-Samorodnitsky-book}.
These models are highly parametrizable, and spectral scale invariance has been demonstrated for 
pulses with a power-law shape \cite{2005-Lowen-book},
as well as for distributed inter-event times \cite{1993-Lowen-PhysRevE.47.992, 2005-Kaulakys-PhysRevE.71.051105, 2009-Silvestri-PhysRevLett.102.014502, 2016-Leibovich-PhysRevE.94.052130},
relaxation rates \cite{2005-Milotti-PhysRevE.72.056701}, and pulse durations \cite{1950-Ziel, 1972-Butz-JStatPhys}. Recently, a power spectral density scaling as $1/f$ at low frequencies has been shown to result from a sequence of nonoverlapping rectangular pulses with a power law distribution of durations or waiting times \cite{2023-Kononovicius-PhysRevE.107.034117}.

In the highly cited paper by van der Ziel \cite{1950-Ziel}, the spectral density of semi-conductor noise $X(t)$ with a probability density of event durations $P_\tau(\tau)$ is constructed as a weighted sum of 
individual spectral densities, each corresponding to a signal with a single relaxation time,
\begin{equation}\label{eq:ziel}
    4\langle{X(t)^2}\rangle\, \int_0^\infty \rmd\tau\,\tau\, P_{\tau}(\tau)/(1+\tau^2\omega^2),
\end{equation} 
where $\omega=2\pi f$ is the angular frequency, and the Lorentzian term $(1+\tau^2\omega^2)^{-1}$ corresponds to the spectral density of a one-sided exponential pulse function.
Importantly, \Eqref{eq:ziel} does not reflect the dynamics of a superposition of uncorrelated pulses with durations distributed continuously over a wide range. 
In such a process the overlapping long-lasting events instigate internal long-range correlations resulting in algebraic tails of the auto-correlation function. As remarked by Butz \cite{1972-Butz-JStatPhys}, the auto-correlation function is then itself $\tau$-dependent, and hence the term $\langle{X(t)^2}\rangle$ should be included in the integration, yielding a spectrum proportional to
\begin{equation}\label{eq:butz}
    \int_0^\infty \rmd\tau\,\tau^2\, P_{\tau}(\tau)/(1+\tau^2\omega^2).
\end{equation}  
For $P_\tau\sim1/\tau^\alpha$, \Eqsref{eq:ziel} and~\eqref{eq:butz} indicate different scaling signatures of the frequency power spectral density: $1/\omega^{2-\alpha}$ and $1/\omega^{3-\alpha}$, respectively. Obtaining $1/f$ noise thus calls for $P_\tau\sim1/\tau$ according to \Eqref{eq:ziel}, and $P_\tau\sim1/\tau^2$ according to \Eqref{eq:butz}.
An empirical $1/f$-like scaling has been demonstrated for an aggregation of random telegraph noise oscillators with individual Lorentzian characteristics \cite{2017-Costanzi-PhysRevLett.119.097201-1oF-superpos-Lorentzians}, speaking to van der Ziel's \Eqref{eq:ziel}.
Such an aggregate is, however, distinct from a single process consisting of superposed, uncorrelated pulses with distributed durations, whose power spectral density was shown by Butz to follow \Eqref{eq:butz}.

The literature reflects the lack of clarity and discernment on the matter.
In Ref.~\onlinecite{1998-Jensen-book}, the $1/\omega^{3-\alpha}$ scaling is explicitly acknowledged alongside the derivation of \Eqref{eq:ziel} and a discussion of the resulting $1/\omega^{2-\alpha}$ scaling. 
The ambiguity is furthered by yet another variant of the spectral density, plausibly dating back to Bernamont \cite{1937-Bernamont},
\begin{equation}\label{eq:bernamot}
    \int_{\taumin}^{\taumax} \rmd\tau\, P_{\tau}(\tau)/(1+\tau^2\omega^2),
\end{equation} 
where $\taumin$ and $\taumax$ are, respectively, the minimum and maximum pulse durations.
Equation~\eqref{eq:bernamot} is mentioned in Refs.~\onlinecite{2020-Fleetwood-chapter-citing-Bernamot} and \onlinecite{2023-Fleetwood-citin-Ziel} as a substantiation of $1/f$ noise originating from $P_\tau\sim1/\tau$.
We remark that integrating \Eqref{eq:bernamot} with $P_\tau = 1/\tau$ yields a logarithmic expression, ${\ln [\taumax^2(1+\taumin^2\omega^2)/(\taumin^2(1+\taumax^2\omega^2))]}\nsim 1/\omega$, which does not display the $1/\omega$ signature even in the limit of an infinitely wide distribution. A uniform $P_\tau$ is required for obtaining the $1/\omega$ scaling from \Eqref{eq:bernamot}.

The main objective of this paper is to present a comprehensive and systematic overview over the scaling characteristics of the frequency power spectral density of a Poisson process with distributed pulse durations, extending the original work by Butz \cite{1972-Butz-JStatPhys} and Milotti \cite{2005-Milotti-PhysRevE.72.056701}. 
Our analysis includes a derivation of a general integral expression for the power spectral density, followed by closed-form expressions obtained for a one-sided exponential pulse function and several power-law pulse duration distributions with various truncations.
The $1/\omega^{3-\alpha}$ scaling of the power spectral density is demonstrated explicitly in the limit of an infinitely broad distribution of durations $P_\tau(\tau)\sim1/\tau^\alpha$ for ${1<\alpha<3}$,
with logarithmic corrections to the frequency scaling exposed for $\alpha=1$ and $\alpha=3$.
The emergence of the $1/\omega^{3-\alpha}$ scaling for bounded regions of scale invariance is confirmed by the compensated spectra of the analytical expressions as well as from realizations of the process. 
This also exposes the finite-size effects associated with the truncation of the underlying pulse duration distributions. 
It is also demonstrated that for propagating fixed-size pulses observed from a single location, the $P_\tau(\tau)\sim1/\tau^2$ distribution of their durations is equivalent to a uniform distribution of their velocities, removing the requirement of power-law input for obtaining a $1/f$ noise signature. 

This paper is organized as follows. 
In \Secref{sec:model} we present the stochastic process describing fluctuations in physical systems as a superposition of uncorrelated pulses. General expressions are derived for the auto-correlation function and the frequency power spectral density for a distribution of pulse durations. 
Power-law distributions with various support truncations are defined in \Secref{sec:distributed-Ptau}, with complementary definitions presented in \Appref{sec:psd-other-Ptau}.
Power-law distributed pulse durations are motivated in \Appref{sec:motiv}, where for an advection equation we demonstrate that the probability density of pulse durations is characterized by power-law scalings when the corresponding pulse propagation velocities follow a uniform or a Gamma distribution. 
In \Secref{sec:regime-argument} we discuss the infinite variance of pulse durations as a prerequisite for long-range correlations, and we establish a relation between the scaling of the frequency power spectral density and the scaling of the underlying distribution of pulse durations.
Closed-form power spectral density expressions demonstrating scale invariance and logarithmic corrections to the frequency scaling are derived in \Secref{sec:PSD}, with complementary derivations presented in \Appref{sec:psd-other-Ptau}. A discussion of the results and conclusions follow in \Secsref{sec:discussion} and~\ref{sec:conclusions}, respectively.

\section{Stochastic Model}\label{sec:model}

In this section we outline the stochastic process describing fluctuations due to a super-position of uncorrelated pulses with a distribution of durations. General expressions for the two lowest-order moments, the auto-correlation function and the power spectral density are derived. Considering a normalized random variable, a general expression for the dimensionless power spectral density is obtained.

\subsection{Super-position of uncorrelated pulses}

Consider a stochastic process given by a superposition of $K$ uncorrelated pulses distributed uniformly in a time interval of duration $T$ \cite{2017-garcia-PhysPlasm}, 
\begin{equation}\label{PhiK_shotnoise}
    \Phi_K(t) = \sum_{k=1}^{K(T)} A_k\phi\left( \frac{t-\tarriv_k}{\taudim_k} \right) .
\end{equation}
Each pulse indexed by $k$ is characterized by an amplitude $A_k$, a duration $\taudim_k$, and an arrival time $\tarriv_k$, the latter following a uniform probability density, $P_\tarriv=1/T$.
The pulse function $\phi(\theta)$ is normalized such that $\int_{-\infty}^{\infty} \rmd\theta\,\abs{\phi(\theta)} = 1$. 
The total number of pulses $K$ arriving in the interval of duration $T$ is a random variable distributed according to a Poisson distribution,
\begin{equation}
    P_K(K;T)=\frac{1}{K!}\left(\frac{T}{\save{\twait}}\right)^K \, \exp\left(-\frac{T}{\save{\twait}}\right) ,
\end{equation}
where $\save{\twait}$ is the average pulse waiting time. In a Poisson process, the waiting times between consecutive pulses are exponentially distributed.
The pulse durations $\taudim_k$ are assumed to be distributed randomly with probability density $P_\taudim(\taudim)$, and an average pulse duration $\langle \taudim \rangle = \int_0^\infty \rmd\taudim\,\taudim\,P_\taudim(\taudim)$. Each of the pulse parameters $A$, $\tarriv$ and $\taudim$ are assumed to be independent and identically distributed.

Given the distribution of pulse amplitudes $P_A(A)$ with a finite mean $\Aave$ and variance $\langle{A^2}\rangle$, the moments and the auto-correlation function of the process defined by \Eqref{PhiK_shotnoise} with exactly $K$ pulses are calculated by averaging over all random variables according to Campbell's theorem \cite{1909-Campbell-ProcCambr, 1944-Rice-MathematicalAO}. 
Subsequently, averaging over the randomly distributed number of pulses $K$ yields the corresponding expressions for the the stationary process \cite{2017-garcia-PhysPlasm}.

\subsection{Mean and variance}

For uniformly distributed pulse arrival times, the conditional mean value of the process with exactly $K$ pulses is given by 
\cite{2017-garcia-PhysPlasm}
\begin{equation}
    \langle{\Phi_K}\rangle=\save{\taudim} \save{A} I_1 K/T,   
\end{equation}
where $I_n$ is the integral of the $n$'th power of the pulse function defined for positive integers $n$ as
\begin{equation}
    I_n = \int_{-\infty}^{\infty} \rmd\theta\,\left[ \phi(\theta) \right]^n .
\end{equation}
With $K(T)$ following a Poisson distribution, the average number of pulses in an interval of duration $T$ is given by $\save{K} = {T}/{\save{\twait}}$.
The mean value of the stationary process is then given by
\begin{equation}\label{Phi-mean}
    \save{\Phi} = \gamma\,I_1\save{A},
\end{equation}
where $\gamma=\save{\taudim}/\save{\twait}$ is the intermittency parameter determining the degree of pulse overlap \cite{2016-garcia-PhysPlasm}.

The second order moment is calculated in a similar manner, averaging over the square of the random variable $\langle{\Phi_K^2}\rangle$.
The variance of the process then follows as \cite{2017-garcia-PhysPlasm}
\begin{equation}
    \langle{\Phi^2}\rangle = \Phiave^2 + \Phirms^2 ,
\end{equation}
where the standard deviation is given by 
\begin{equation}
    \Phirms^2 = \gamma\,I_2\langle{A^2}\rangle .
\end{equation}
The absolute level of fluctuations is thus proportional to the degree of pulse overlap described by the parameter $\gamma$. Long pulse durations and short waiting times both contribute to the increase in the value of $\gamma$ and the mean value of the process given by \Eqref{Phi-mean}.
Conversely, the relative level of fluctuations of the process with non-zero mean $\save{\Phi}$ is described by
\begin{equation}
    \frac{\Phirms^2}{\langle{\Phi}^2\rangle} = \frac{1}{\gamma}\, \frac{I_2}{I_1^2}\, \frac{\langle{A^2}\rangle}{\save{A}^2},
\end{equation}
which is large when there is a clear separation of pulses with relatively short durations, corresponding to $\gamma\ll1$.
The moments and the characteristic function of the process given by \Eqref{PhiK_shotnoise} are independent of the distribution of pulse durations \cite{2017-garcia-PhysPlasm}.

\subsection{Auto-correlation function}\label{sec:acf}

The auto-correlation function of the process $\Phi_K(t)$ for a time lag $r$ is calculated as
\begin{align}
    R_{\Phi_K}(r) &= \langle{\Phi_K(t)\Phi_K(t+r)}\rangle .\label{autocorrK}
\end{align}
For a normalized, non-negative pulse function $\phi(\theta)$, averaging over the randomly distributed number of pulses $K$ gives the auto-correlation function for the stationary process \cite{2017-garcia-PhysPlasm},
\begin{equation}
    R_\Phi(r) = {\Phiave}^2 + \Phirms^2 \frac{1}{\save{\taudim}} \int_0^\infty \rmd \taudim\, \taudim\, P_\taudim(\taudim)\, \rho_\phi(r/\taudim),\label{autocorrPhi}
\end{equation}
where
\begin{equation}\label{rhophi}
    \rho_\phi(\theta) = \frac{1}{I_2}\int_{-\infty}^{\infty} \rmd\chi\,\phi(\chi)\phi(\chi+\theta)
\end{equation}
is a normalized auto-correlation function of the pulse function.

\subsection{Power spectral density}\label{sec:psd}

The power spectral density is given by the Fourier transform of the auto-correlation function in \Eqref{autocorrPhi} \cite{2017-garcia-PhysPlasm},
\begin{align}\label{psdphi}
    \Omegadim_{{\Phi}}(\omegadim) &= \int_{-\infty}^{\infty} \rmd r\,R_\Phi(r)\,\exp(-i\omegadim r)\nonumber\\
    &= 2\pi\save{\Phi}^2\delta(\omegadim) + \Phirms^2 \frac{1}{\save{\taudim}}\int_0^\infty \rmd\taudim\,\taudim^2 P_{\taudim}(\taudim) \Varrho_\phi(\taudim\omegadim),
\end{align}
where $\omegadim=2\pi f$ is the dimensional angular frequency and 
\begin{equation}\label{varrhophi}
    \Varrho_\phi(\vartheta) = \int_{-\infty}^\infty \rmd\theta\,\rho_\phi(\theta)\exp(-i\vartheta\theta) = \frac{1}{I_2}\,\abs{\varphi}^2(\vartheta) 
\end{equation}
is the Fourier transform of the normalized auto-correlation of the pulse function given by \Eqref{rhophi} and
\begin{equation}\label{varphi}
    \varphi(\vartheta) = \int_{-\infty}^\infty \rmd\theta\,\phi(\theta)\exp(-i\vartheta\theta) 
\end{equation}
is the Fourier transform of the pulse function $\phi(\theta)$.
The power spectral density given by \Eqref{psdphi} is independent of the distribution of pulse amplitudes $P_A(A)$.
It is derived for uniformly distributed pulse arrival times, and under the assumption that the two lowest-order moments of the process are finite.

\subsection{Normalization}\label{sec:nodimPSD}

Consider a rescaled random variable with zero mean and unit standard deviation,
\begin{equation}\label{Phitilde}
    \Phiwt = \frac{\Phi-\Phiave}{\Phirms} .
\end{equation}
The auto-correlation function for the normalized process follows as
\begin{equation} \label{autocorrPhiTilde}
    R_{\Phiwt}(r) = \frac{1}{\save{\taudim}}\int_0^\infty \rmd\taudim\,\taudim P_\taudim(\taudim) \rho_\phi(r/\taudim),
\end{equation}
and the power spectral density, given by the Fourier transform of \Eqref{autocorrPhiTilde}, becomes
\begin{equation}\label{psdPhiTilde}
    \Omegadim_{\wt{\Phi}}(\omegadim) = \frac{1}{\save{\taudim}}\int_0^\infty \rmd\taudim\,\taudim^2 P_{\taudim}(\taudim) \Varrho_\phi(\taudim\omegadim).
\end{equation}
It should be noted that for the normalized variable, the auto-correlation function and the power spectral density are independent of the intermittency parameter $\gamma$. They are defined in terms of the distribution of pulse durations $P_\taudim(\taudim)$ and the transformations of the pulse function, $\rho_\phi(\theta)$ and $\Varrho_\phi(\vartheta)$, given respectively by \Eqsref{rhophi} and~\eqref{varrhophi}. 

Normalizing $\taudim$ and $\omegadim$ by the mean value $\save{\taudim}$, the dimensionless pulse duration and the angular frequency are introduced respectively as
\begin{subequations}
\begin{align}
    \tau&={\taudim}/{\save{\taudim}},\label{dimless-tau}\\
    \omega&=\omegadim\,\save{\taudim}\label{dimless-omega}.
\end{align}   
\end{subequations}
The normalized probability density function of the dimensionless pulse durations is then given by
\begin{equation}\label{Pdimless-dim}
    P_{\tau}(\tau) = P_{{\taudim}/{\save{\taudim}}}({\taudim}/{\save{\taudim}}) = \save{\taudim}\,P_\taudim(\taudim),
\end{equation} 
and the power spectral density in \Eqref{psdPhiTilde} becomes 
\begin{equation}\label{psd-dimensional-mean}
    \Omegadim_{\wt{\Phi}}(\omega/\save{\taudim}) = \save{\taudim} \int_0^\infty \rmd\tau\,\tau^2 P_{\tau}(\tau) \Varrho_\phi(\tau\omega).
\end{equation}
The dimensionless power spectral density defined as $\Omega_{\wt{\Phi}}(\omega)=\Omegadim_{\wt{\Phi}}(\omega/\save{\taudim})/\save{\taudim}$ then follows as
\begin{equation}\label{psdphifullydimless}
    \Omega_{\wt{\Phi}}(\omega)=
    \int_0^\infty \rmd\tau\,\tau^2 P_{\tau}(\tau) \Varrho_\phi(\tau\omega) ,
\end{equation}
for a distribution of pulse durations $P_\tau(\tau)$.
In the following, only the dimensionless power spectral density form in \Eqref{psdphifullydimless} is considered.

\subsection{Spectral scaling properties}

In the special case of a constant pulse duration, the power spectral density in \Eqref{psdphifullydimless} is fully determined by the power spectrum of the pulse function.
For example, for a one-sided exponential pulse function defined in \Eqsref{phirhovar-1exp} the power spectral density is given by a Lorentzian function,
\begin{equation}\label{psd-exp}
    \Omega_{\Phiwt}(\omega) = \frac{2}{1+\omega^2} ,
\end{equation}
which has a break point at $\omega=1$, and two asymptotic scaling regimes for the limiting frequencies: the constant value $2$ in the limit $\omega\rightarrow0$, and a $2/\omega^2$ tail in the limit $\omega\rightarrow\infty$. 
There is no frequency range of scale invariance characterized by exponents $0<\beta<2$.

Maintaining a constant pulse duration and introducing a power-law pulse function was shown to result in spectral scale invariance in a range of scaling exponents and for certain truncations of the power-law function \cite{2005-Lowen-book}.
In the following, we demonstrate that the frequency power spectral density in \Eqref{psdphifullydimless} manifests self-similar scaling for certain power-law distributions of pulse durations irrespective of the pulse function.

\section{Power-law distributed pulse durations}\label{sec:distributed-Ptau}%

In this contribution, scale invariance is introduced into the dimensionless power spectral density in \Eqref{psdphifullydimless} through the distribution of pulse durations.
Several variants of power-law distributions are considered in this section. The conditions under which these distributions result in long-range dependence are then analyzed in \Secref{sec:regime-argument}.
The origins of power-law distributions of pulse durations are motivated for a superposition of propagating pulses in \Appref{sec:motiv}.

The probability densities considered here are normalized such that
\begin{equation}\label{normalized_Ptau}
    \int_0^\infty \rmd\tau\,P_\tau(\tau)=1,
\end{equation}
with the mean value following from \Eqref{dimless-tau},
\begin{equation}\label{normalized_taumean} 
    \save{\tau}=\int_{0}^{\infty} \rmd\tau\,\tau\, P_\tau(\tau) =1.    
\end{equation}
A finite, nondivergent mean is required for the stationarity of the stochastic process given by \Eqref{PhiK_shotnoise} and for the normalization of the power spectral density in \Eqref{psdphifullydimless}.

\subsection{Variants of power-law distributions}\label{sec:variants-Pareto}

Consider a general form of Pareto distribution with an exponent $\alpha \geq 0$,
\begin{equation}\label{gpareto}
    P_\tau(\tau;\alpha,\taumin,\taumax) =
    \begin{cases}
        \displaystyle \eta\,\tau^{-\alpha}  & \text{if }\tau\in\left[\taumin,\taumax\right] , \\
        0 & \text{otherwise},
    \end{cases}
\end{equation}
where $\eta$ is a normalization factor and $\taumin < \taumax$ are the support boundaries.
Following the classification proposed in Ref.~\onlinecite{2005-Lowen-book} for the power-law pulse function we distinguish four variants of Pareto distribution with different support truncations: 
unbounded $\tau \in \left(0, \infty\right)$, upper-truncated $\tau \in \left(0, \taumax \right]$, lower-truncated $\tau \in \left[\taumin, \infty\right)$, and bounded $\tau \in \left[\taumin, \taumax \right]$.
The configuration of the support boundaries and the normalization conditions in \Eqsref{normalized_Ptau} and~\eqref{normalized_taumean} determine the expressions for $\eta$, $\taumin$ and $\taumax$ as functions of $\alpha$.

The bounded Pareto distribution will be shown most relevant for modelling long-range correlations with finite-size effects.
Its definition follows in \Secref{sec:Pareto-bounded}, 
while the upper-truncated and the lower-truncated Pareto distributions are defined in \Appsref{sec:Pareto-upper-truncated} and~\ref{sec:Pareto-standard}, respectively.
The unbounded Pareto distribution is not well defined and cannot be normalized to comply with \Eqref{normalized_Ptau}.
Instead, an inverse Gamma distribution is considered,
as it is well defined for an unbounded support $\tau \in \left(0, \infty\right)$ due to a smooth exponential cutoff for small values of $\tau$.
The inverse Gamma distribution of pulse durations is motivated in \Appref{sec:gamma-velocities} and defined in \Appref{sec:inverse-Gamma-distr}.

\subsection{Bounded Pareto distribution}\label{sec:Pareto-bounded}

The Pareto distribution defined by \Eqref{gpareto} with a support truncated on both sides $\tau \in \left[\taumin, \taumax \right]$ is well defined for any $\alpha \geq 0$. Applying the condition \eqref{normalized_Ptau} yields the following normalization factor with its limit
\begin{subequations}\label{etas-bPareto}
    \begin{align}
        &\eta (\alpha, \taumin, \taumax) = \frac{(\alpha-1) (\taumax \taumin)^\alpha}{\taumax^\alpha \taumin - \taumax \taumin^\alpha},\label{eta-bPareto}\\
        &\lim_{\alpha\rightarrow1}  \eta(\alpha, \taumin, \taumax) = \left(\ln\frac{\taumax}{\taumin}\right)^{-1}.
    \end{align}
\end{subequations}
The bounded Pareto distribution has a total of three parameters $\{\alpha, \taumin, \taumax\}$ subject to the conditions \eqref{normalized_Ptau} and~\eqref{normalized_taumean}. 
We thus reduce the number of parameters by introducing the ratio of the two support boundaries as a new dimensionless parameter,
\begin{equation} \label{delta}
    \Delta = \frac{\taumax}{\taumin} ,
\end{equation}
where it follows that $\Delta>1$. The parameter $\Delta$ describes the logarithmic width of the bounded Pareto distribution, where $\log_{10}\Delta$ gives the number of logarithmic decades spanned by the support.

The normalization factor and its limit given by \Eqsref{etas-bPareto} are expressed in terms of $\alpha$ and $\Delta$ as
\begin{subequations}\label{etas-bPareto-Delta}
    \begin{align}
        &\eta(\alpha,\Delta) = \frac{(\alpha-1)}{1 - \Delta^{1-\alpha}} \taumin^{\alpha-1},\label{eta-bPareto-Delta}\\
        &\lim_{\alpha\rightarrow1} \eta(\alpha,\Delta) = \frac{1}{\ln\Delta} .
    \end{align}
\end{subequations}
The resulting mean of the bounded Pareto distribution normalized according to the condition \eqref{normalized_taumean} yields the two support boundaries with their respective limits,
\begin{subequations}\label{eq:taus-Delta}
\begin{align}
    &\taumin(\alpha,\Delta) = \frac{(\alpha-2) (1 - \Delta^{1 - \alpha})}{(\alpha-1) (1 - \Delta^{2 - \alpha})} ,\\
    &\lim_{\alpha\rightarrow1}  \taumin(\alpha,\Delta) = \frac{\ln\Delta}{\Delta-1},\\
    &\lim_{\alpha\rightarrow2}  \taumin(\alpha,\Delta) = \frac{\Delta-1}{\Delta\,\ln\Delta},\\
    &\taumax(\alpha,\Delta) = \Delta\, \taumin(\alpha,\Delta).\label{taumax-bPareto-Delta}
\end{align}
\end{subequations}
The bounded Pareto distribution is thus parametrized as $P_{\tau}(\tau;\alpha,\Delta)$. This definition is aligned with \Eqref{gpareto-b} motivated in \Appref{sec:motiv-pareto} for the dimensional variable. 
The variance of the pulse duration is then given by
\begin{equation}
    \langle{\tau^2}\rangle(\alpha,\Delta) = \frac{(\alpha-2)(1 - \Delta^{3 - \alpha})}{(\alpha-3)(1 - \Delta^{2 - \alpha})} \taumin ,
\end{equation}
with the following finite limits
\begin{subequations}\label{var-bPareto}
    \begin{align}
        \lim_{\alpha\rightarrow1} \save{\tau^2}(\alpha,\Delta)  & = \frac{(\Delta+1)\ln \Delta}{2(\Delta-1)},\\
        \lim_{\alpha\rightarrow2} \save{\tau^2}(\alpha,\Delta)  & = \frac{(\Delta-1)^2}{\Delta\, \ln^2 \Delta},\\
        \lim_{\alpha\rightarrow3} \save{\tau^2}(\alpha,\Delta)  & = \frac{(\Delta+1)\ln \Delta}{2(\Delta-1)}.
    \end{align}
\end{subequations}
The variance and the parameters of the distribution determine the long-range dependence characteristics of the associated process, as will be discussed in \Secref{sec:regime-argument}.

\begin{figure}
    \includegraphics[width=0.5\textwidth]{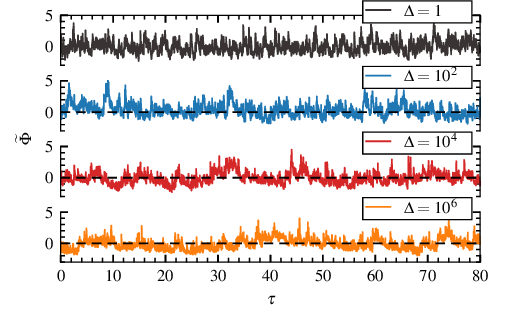 }
    \caption{Normalized realizations of the stochastic process for a one-sided exponential pulse function, pulse overlap parameter $\gamma=10$, and a bounded Pareto distribution of pulse durations with exponent $\alpha=2$. Consecutive realizations are obtained for different distribution widths $\Delta$.}
    \label{fig:TS-alpha-delta}
\end{figure}

\begin{figure}
    \includegraphics{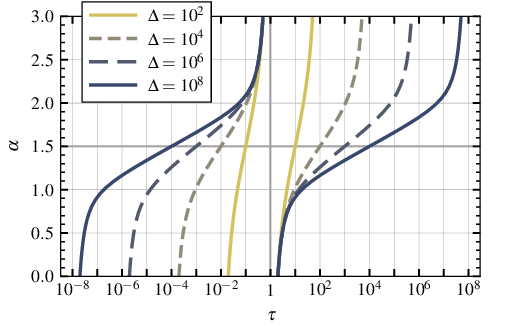}
    \caption{
        Range of normalized pulse durations $\tau$ for a bounded Pareto distribution and various width parameters $\Delta$. Each pair of corresponding curves marks the boundaries $\taumin$ and $\taumax$ for a given distribution width $\Delta$ and with the power-law exponent $\alpha$ varying in the range $\alpha\in[0,3]$. Pulse durations are normalized with the mean value.}
    \label{fig:tau-range}
\end{figure}

Figure \ref{fig:TS-alpha-delta} shows realizations of the process defined by \Eqref{PhiK_shotnoise} for the bounded Pareto distribution of pulse durations for $\alpha=2$ and different values of $\Delta$.
The top curve for $\Delta=1$ corresponds to a process with a constant pulse duration, characterized by a Lorentzian power spectral density given by \Eqref{psd-exp}.
Extending the width of the distribution entails an increase in the upper bound in \Eqref{taumax-bPareto-Delta}.
The resulting introduction of pulses with relatively long durations 
is manifested as low-frequency fluctuations, which in \Figref{fig:TS-alpha-delta} become gradually more pronounced as the value of $\Delta$ is increased.
Importantly, the extent to which broadening of the distribution width affects the upper bound of the distribution and the long-range dependence characteristics of the process depends on the value of $\alpha$. This is demonstrated in \Figref{fig:tau-range} and further addressed in \Secref{sec:divergence}.

\section{Conditions for long-range dependence}\label{sec:regime-argument}

The scale invariance of the frequency power spectral density is related to the self-similarity properties of the underlying distribution of pulse durations. 
In this section, we consider the distributions defined in \Secref{sec:distributed-Ptau} and \Appref{sec:psd-other-Ptau}, and we identify the parameter ranges at which long-range dependence is expected to emerge.
The results are summarized in Table~\ref{tab:alpha-regimes-expected-scaling}.

\subsection{Infinite variance}

Long-range correlations are typically well represented by processes with infinite variance. 
For the lower-truncated Pareto and the inverse Gamma distributions 
the dimensional pulse duration $\taudim$ is characterized by a finite mean value and an infinite variance for $2 < \alpha \leq 3$. For $\alpha\leq2$ the mean value is infinite, preventing the normalization of the power spectral density $\Omega_{\wt{\Phi}}$ given by \Eqref{psdphifullydimless}.
These findings are reflected in the last two columns of Table~\ref{tab:alpha-regimes-expected-scaling}.

For the upper-truncated Pareto and the bounded Pareto distributions the variance of the pulse durations is finite for all values of $\alpha$.
However, this does not preclude long-range dependence resulting from these finite-support distributions. In the following we discuss further distribution aspects that may promote spectral scale invariance and long-range correlations.

\subsection{Long pulse durations}\label{sec:divergence}

Pulses with long durations obviously contribute to long-range dependence. Allowing arbitrarily long pulse durations requires a divergence of the upper bound, $\taumax\rightarrow\infty$. 
For the upper-truncated Pareto distribution, the upper bound given by \Eqref{upper-tr-Pareto-last} diverges towards infinity only in the limit $\alpha\rightarrow1^-$. In this limit the distribution itself is not well defined. The power spectral density of a process with an upper-truncated Pareto distribution of pulse durations is thus not expected to manifest scale invariance. This is indicated in the first column of Table~\ref{tab:alpha-regimes-expected-scaling}.

In the case of the bounded Pareto distribution, both support boundaries are modified simultaneously and indirectly, through the parameter $\Delta$ given by \Eqref{delta}. Changing the value of $\Delta$ affects the behavior of $\taumin$ and $\taumax$ differently depending on the value of $\alpha$.
In particular, in the limit $\Delta\rightarrow\infty$
the lower and the upper bounds given by \Eqsref{eq:taus-Delta}, as well as the normalization factor given by \Eqref{etas-bPareto-Delta}, have the following limits
\begin{subequations}\label{tau-limit-delta-infinity}
    \begin{align}
    &\lim_{\Delta\rightarrow\infty}\taumin(\alpha,\Delta) = 
    \begin{cases}
        0  & \text{if }\alpha\leq2, \\
        \frac{\alpha-2}{\alpha-1} & \text{if }\alpha>2,
    \end{cases}\label{tmin-limit-delta-infinity}\\
    &\lim_{\Delta\rightarrow\infty}\taumax(\alpha,\Delta)=
    \begin{cases}
        \frac{\alpha-2}{\alpha-1} & \text{if }\alpha<1, \\
        \infty & \text{if }\alpha\geq 1,
    \end{cases}\label{tmax-limit-delta-infinity}\\
    &\lim_{\Delta\rightarrow\infty}\eta(\alpha,\Delta)=
    \begin{cases}
        -\frac{(\alpha-2)^{\alpha-1}}{(\alpha-1)^{\alpha-2}} & \text{if }\alpha < 1, \\
        0 & \text{if }1\leq\alpha\leq2,\\
        \frac{(\alpha-2)^{\alpha-1}}{(\alpha-1)^{\alpha-2}}  & \text{if }\alpha > 2.\label{eta-limit-delta-infinity}
    \end{cases}
\end{align}
\end{subequations}
Equations~\eqref{tau-limit-delta-infinity} indicate that in the limit $\Delta\rightarrow\infty$ the bounded Pareto distribution approaches the upper-truncated Pareto distribution when $\alpha<1$, and approaches the lower-truncated Pareto distribution when $\alpha>2$. In the range $1\leq\alpha\leq2$ the lower and the upper bounds approach zero and infinity respectively, while the normalization factor vanishes, ensuring that the normalization condition \eqref{normalized_taumean} is met.

The placement of the support bounds with respect to the mean value is demonstrated in \Figref{fig:tau-range} for various values of the width parameter $\Delta$ and for scaling exponents $\alpha\in[0,3]$.
For ${\alpha<1}$ the increase in $\Delta$ implies the decrease of the lower bound $\taumin$ with respect to the mean value, while the upper bound $\taumax$ remains at the order of magnitude of the mean value. The opposite is true for $\alpha>2$, where $\taumax$ increases with respect to the mean value while $\taumin$ remains close to the mean value. 
Figure~\ref{fig:tau-range} shows a smooth transition between these two regimes. At $\alpha=3/2$ both $\taumin$ and $\taumax$ are at equal logarithmic distance from the mean value.

Considering the divergence of the upper bound as the driver of long-range dependence, \Eqref{tmax-limit-delta-infinity} suggests that $\alpha\geq1$ is necessary for arbitrarily long pulse durations to emerge from a sufficiently wide bounded Pareto distribution. As the value of $\Delta$ approaches infinity, the pulse duration variance given by \Eqref{var-bPareto} has the following limits
\begin{equation}
    \lim_{\Delta\rightarrow\infty} \save{\tau^2}(\alpha,\Delta) =
    \begin{cases}
         \infty & \text{if }1\leq\alpha\leq3,\\
         \frac{(\alpha-2)^2}{(\alpha-3)(\alpha-1)}& \text{otherwise} .\label{var-limit-delta-infinity}
    \end{cases}
\end{equation}
Long-range dependence is thus expected in the exponent range $1\leq\alpha\leq3$.

As the power-law exponent increases beyond $\alpha=3$, the bounded Pareto distribution narrows and assumes the shape of a degenerate distribution corresponding to constant pulse duration. Conversely, in the limit $\alpha\rightarrow0$ the bounded Pareto distribution reduces to a uniform distribution, with no long-range correlations involved \cite{2017-garcia-PhysPlasm}.

\begin{table}
    \centering
    \caption{Existence of the probability density function and the two lowest-order moments for pulse durations at different values of exponent $\alpha$. Colored shading indicates the anticipated self-similar scaling of the power spectral density due to infinite variance.}
    \label{tab:alpha-regimes-expected-scaling}
    \begin{ruledtabular}
    \begin{tabular}{|c||@{}c@{}|@{}c@{}|@{}c@{}|@{}c@{}|}
        & \multicolumn{3}{c|}{ \shortstack{ $P_\tau$ Pareto \\[0pt]}} 
        & \shortstack{ $P_\tau$ inverse\\ Gamma}\\\cline{2-5} 
        & $\;\;\tau \in \left(0, \taumax \right]\;\;$
        & $\;\;\tau \in \left[\taumin, \taumax \right]\;\;$
        & $\;\;\tau \in \left[\taumin, \infty\right)\;\;$ 
        & $\;\;\tau\in(0,\infty)\;\;$
        \\ \colrule\colrule
        $0<\alpha<1$
        & \ccNoScal \footnotesize $\save{\tau^2}<\infty$
        & \ccNoScal \footnotesize $ \displaystyle \lim_{\Delta\rightarrow\infty}\save{\tau^2}<\infty$
        & \multicolumn{2}{@{}c@{}|}{\ccNoPdf \footnotesize no $P_\taudim$} 
        \\ \cline{1-3} 
        $\alpha=1$
        & \ccNoPdf \footnotesize no $P_\taudim$
        & \ccYesScal 
        & \multicolumn{2}{@{}c@{}|}{\ccNoPdf } 
        \\ \cline{1-1}\cline{4-5} 
        $1<\alpha\leq2$
        & \ccNoPdf 
        & \ccYesScal \footnotesize$\displaystyle \lim_{\Delta\rightarrow\infty}\taumax=\infty$
        & \multicolumn{2}{@{}c@{}|}{\ccNosAve \footnotesize $\save{\taudim}=\infty$} 
        \\\cline{1-1} \cline{4-5}
        $2<\alpha<3$
        & \ccNoPdf 
        & \ccYesScal \footnotesize$ \displaystyle \;\;\lim_{\Delta\rightarrow\infty}\save{\tau^2}=\infty\;\;$
        & \multicolumn{2}{@{}c@{}|}{\ccYesScal \footnotesize $\save{\tau^2}=\infty$}
        \\\cline{1-1}
        $\alpha=3$
        & \ccNoPdf 
        & \ccYesScal 
        & \multicolumn{2}{@{}c@{}|}{\ccYesScal }
        \\\cline{1-1}\cline{3-5} 
        $\alpha>3$ 
        & \ccNoPdf 
        & \ccNoScal \footnotesize$ \displaystyle \lim_{\Delta\rightarrow\infty}\save{\tau^2}<\infty$
        & \multicolumn{2}{@{}c@{}|}{\ccNoScal \footnotesize$\save{\tau^2}<\infty$\normalsize} 
    \end{tabular}
    \end{ruledtabular}
\end{table}

Long-range dependence is anticipated for the distributions which allow for arbitrarily long pulse durations, and in the range of exponents $\alpha$ where the pulse duration variance is infinite.
The bounded Pareto distribution represented in the second column of Table~\ref{tab:alpha-regimes-expected-scaling} satisfies both conditions for the widest range of exponents $\alpha$.
These predictions are verified in \Secref{sec:PSD}, where the corresponding Table~\ref{tab:alpha-regimes-PSD-scaling} summarizes the signatures of scale invariance demonstrated for the analytical power spectral density expressions.

\begin{figure*}
    \includegraphics[width=0.5\textwidth]{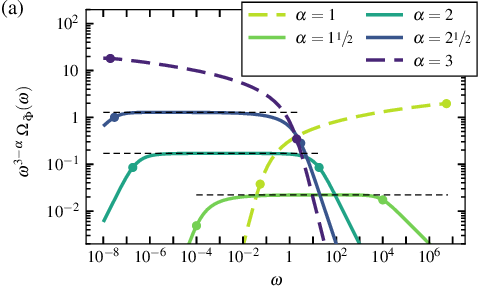}%
    \includegraphics[width=0.5\textwidth]{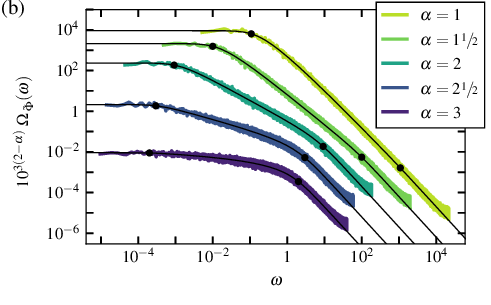}
    \caption{
        Power spectral densities for a super-position of one-sided exponential pulses with bounded-Pareto distributed durations. 
        Filled circles mark the cutoff frequencies $\omega\taumax=1$ and $\omega\taumin=1$.
        The presented values of $\alpha$ span the entire exponent range associated with spectral scale invariance obtained for the bounded Pareto distribution.
        (a) Compensated analytical power spectral densities $\omega^{3-\alpha}\,\Omega_{\wt{\Phi}}(\omega; \alpha)$ with $\Omega_{\wt{\Phi}}$ given by \Eqref{eq:psd-bounded} for $\Delta=10^8$.
        The horizontal dashed black lines mark the inverse of the $\omega$-independent factors in \Eqsref{psdlim-alpha1-5}--\eqref{psdlim-alpha2-5}.
        The regions where the dashed black lines overlap with the solid colored lines indicate constant compensated spectra, and hence denote the frequency ranges where frequency scaling $1/\omega^{3-\alpha}$ is manifested. 
        The compensated spectra plotted with thick dashed colored lines are the ones where the logarithmic corrections to the frequency scaling apply, according to \Eqsref{psdlim-alpha1} and~\eqref{psdlim-alpha3}. 
        (b) Empirical power spectral densities for realizations of the process with $\Delta=10^4$. 
        The respective analytical power spectral densities are plotted with solid black curves. 
        Vertical shifting by $\alpha$-dependent factors is applied to avoid overlapping.
        }
    \label{fig:psd-bounded-1exp}
\end{figure*}

\subsection{Scaling exponents relation}\label{sec:scaling-rel}

In the following, we explore the relation between the exponent $\alpha$ of the general Pareto distribution in \Eqref{gpareto} and the scaling exponent of the power spectral density given by \Eqref{psdphifullydimless}. 
The general Pareto distribution displays scale invariance in the unbounded limit
\begin{equation}
    \lim_{\substack{\taumin\rightarrow0 \\ \taumax\rightarrow\infty}} P_\tau(\constfactor\tau) = \lim_{\substack{\taumin\rightarrow0 \\ \taumax\rightarrow\infty}} \constfactor^{-\alpha}P_\tau(\tau),\label{Ptau_scaling}
\end{equation}
where $\kappa$ is a scaling factor. Similarly, rescaling the frequency of the power spectral density by a factor $\constfactor$ gives
\begin{equation}\label{rescaled-psd}
    \Omega_{\wt{\Phi}}(\constfactor\omega) = \int_{0}^{\infty} \rmd\tau\,\tau^2 P_\tau(\tau)\Varrho_\phi(\constfactor\tau\omega).
\end{equation}
Inserting the general form of the Pareto distribution into \Eqref{rescaled-psd}, 
making a change of variable $\tau'=\constfactor\tau$, and applying the scaling relation in \Eqref{Ptau_scaling} gives
\begin{align}\label{psdphiscaled}
    \Omega_{\wt{\Phi}}(\constfactor\omega) 
    &= \constfactor^{\alpha-3} \int_{0}^{\infty} \rmd\tau'\,{\tau'}^2\,  P_\tau(\tau')\Varrho_\phi(\tau'\omega).
\end{align}
It follows from \Eqsref{rescaled-psd} and~\eqref{psdphiscaled} that in the limit of an infinitely broad distribution of pulse durations, where \Eqref{Ptau_scaling} applies, the frequency power spectral density displays scale invariance \cite{2023-Korzeniowska-manus-1},
\begin{align}\label{psdscaling}
    \lim_{\substack{\taumin\rightarrow0 \\ \taumax\rightarrow\infty}} \Omega_{\wt{\Phi}}(\constfactor\omega) 
    &= \lim_{\substack{\taumin\rightarrow0 \\ \taumax\rightarrow\infty}} \constfactor^{\alpha-3}\,\Omega_{\wt{\Phi}}(\omega) .
\end{align}
Equation~\eqref{psdscaling} suggests that the power spectral density itself has a power-law signature $\Omega_{\wt{\Phi}}(\omega) \sim 1/\omega^{\beta}$ with
\begin{equation}\label{eq:beta}
    \beta(\alpha)=3-\alpha.
\end{equation}
If $\Omega_{\wt{\Phi}}(\omega)$ displays scale invariance, then it follows from \Secref{sec:divergence} that long-range correlations are expected in the limit $\Delta\rightarrow\infty$ and within the range of Pareto exponents $1\leq\alpha\leq3$. According to \Eqref{eq:beta}, this corresponds to the range of frequency scaling exponents $0\leq\beta\leq2$, where $\beta=0$, $\beta=1$ and $\beta=2$ are exponents characteristic of white, pink, and Brownian noises, respectively.
The frequency range in which bounded spectral scale invariance is anticipated is determined by the range of the underlying pulse durations, with the delimiting angular frequencies given by $1/\taumax$ and $1/\taumin$.
Outside this range the shape of the power spectral density is determined by the power spectral density of the pulse function, $\Varrho_\phi(\tau\omega)$,
resulting in a broken power law with break points at the cutoff frequencies.

\section{Scale invariance of power spectra}\label{sec:PSD}

In the following, we apply the pulse duration distributions described in \Secref{sec:distributed-Ptau} to the general expression for the frequency power spectral density in \Eqref{psdphifullydimless}. We derive explicit, closed-form expression for the power spectral density $\Omega_{\wt{\Phi}}(\omega; \alpha, \Delta)$ parametrized by the scaling exponent $\alpha$ and the distribution width $\Delta$. We expose analytically the universal $1/\omega^{3-\alpha}$ signature emerging for the parameter regimes discussed in \Secref{sec:regime-argument},
and we demonstrate the finite-size effects associated with the truncations of the underlying distributions.

\subsection{Pulse function}

The power spectral density formulas presented in this section are derived for a one-sided exponential pulse function $\phi$ defined below together with its auto-correlation function $\rho_\phi$, and the respective Lorentzian power spectral density $\Varrho_\phi$ \cite{2017-garcia-PhysPlasm},
\begin{subequations}\label{phirhovar-1exp}
\begin{align}
    \phi(\theta) &= \Theta(\theta)\exp(-\theta),\label{phi-exp}\\ 
    \rho_\phi(\theta) &= \exp(-\abs{\theta}),\label{rhophi-exp}\\
    \Varrho_\phi(\vartheta) &= \frac{2}{1+\vartheta^2},\label{varrhophi-exp}
\end{align}
\end{subequations}
where $\Theta$ is the unit step function
\begin{equation}
    \Theta(\theta) = 
    \begin{cases}
        1 & \text{if } \theta \geq 0,\\
        0 & \text{if } \theta < 0.
    \end{cases}
\end{equation}
The pulse function and its power spectral density are respectively normalized as 
\begin{subequations}
\begin{align}
    \int_{-\infty}^\infty \rmd\theta\,\abs{\phi(\theta)} &= 1,\\
    \int_{-\infty}^\infty \rmd\vartheta\,\Varrho_\phi(\vartheta) &= 2\pi,
\end{align}
\end{subequations}
resulting in the following normalization of the power spectral density given by \Eqref{psdphifullydimless},
\begin{equation}\label{psd-integral-2pi}
    \int_{-\infty}^\infty \rmd\omega\,\Omega_{\wt{\Phi}}(\omega) = 2\pi.
\end{equation}
These normalization constraints are satisfied for all the power spectral densities considered in the following.
The presence of spectral scale invariance for other pulse functions is discussed in \Secref{sec:psd-scling-other-pulse-forms}.

\subsection{Power spectral density for the bounded Pareto distribution of pulse durations} \label{sec:analytic-PSD}

Applying the bounded Pareto distribution given by \Eqsref{gpareto} and~\eqref{etas-bPareto-Delta}--\eqref{eq:taus-Delta} to the power spectral density given by \Eqref{psdphifullydimless} we obtain \cite{2023-Korzeniowska-manus-1}
\begin{multline} \label{eq:psd-bounded}
    \Omega_{\wt{\Phi}}(\omega;\alpha,\Delta) = \\
    \begin{cases}
        \frac{1}{\ln\Delta\;\omega^2}
        \ln\left(\frac{(\Delta-1)^2 + (\Delta\, \ln\Delta\; \omega)^2 }{(\Delta-1)^2 + (\ln\Delta\; \omega)^2}\right) & \text{if }\alpha=1, \\
        \frac{2} {\ln\Delta\;\omega }
        \left[\arctan\left(\frac{(\Delta - 1)\omega}{\ln\Delta}\right) 
        - \arctan\left(\frac{(\Delta - 1)\omega} {\Delta \ln\Delta}\right)\right] & \text{if } \alpha = 2, \\
        \frac{2}{(\Delta^\alpha-\Delta)\omega^2}\,\left[ 
        \Delta^\alpha\;_2F_1\left( 1,\frac{\alpha-1}{2},\frac{\alpha+1}{2};-\inv{\taumin^2\omega^2} \right) \right.  \\ 
        \qquad\qquad\quad \left. - \Delta\;_2F_1\left(1,\frac{\alpha-1}{2},\frac{\alpha+1}{2};-\inv{\taumax^2\omega^2} \right) \right] & \text{otherwise},
    \end{cases}
\end{multline}
where $\taumin$ and $\taumax$ are given by \Eqsref{eq:taus-Delta}.
An explicit power spectral density expression equivalent to \Eqref{eq:psd-bounded} was presented in Ref.~\onlinecite{2005-Milotti-PhysRevE.72.056701} for a Poisson process with distributed pulse decay rates.

Considering the compensated spectra in the limit of an infinitely broad distribution of pulse durations for various exponents reveals power law scalings \cite{2023-Korzeniowska-manus-1},
\begin{subequations}\label{psdlim}
    \begin{align}
        &\lim_{\Delta\rightarrow\infty} \Omega_{\wt{\Phi}}(\omega;1,\Delta) 
        \;\; \omega^2 
        \;\; \frac{\ln{\Delta}}{\ln\left(\omega^2\ln^2{\Delta} \right)}
        \; = 1 ,\label{psdlim-alpha1}
        \\
        &\lim_{\Delta\rightarrow\infty} \Omega_{\wt{\Phi}}\left(\omega;\sfrac{3}{2}, \Delta\right)
        \;\; \abs{\omega}^{3/2}
        \;\; \frac{\sqrt{2}(\sqrt{\Delta}-1)}{\pi\sqrt[4]{\Delta}}
        \; = 1 ,\label{psdlim-alpha1-5}
        \\
        &\lim_{\Delta\rightarrow\infty} \Omega_{\wt{\Phi}}(\omega;2,\Delta)
        \;\; \abs{\omega}
        \;\; \frac{\ln{\Delta}}{\pi}
        \; = 1 ,\label{psdlim-alpha2}
        \\
        &\lim_{\Delta\rightarrow\infty} \Omega_{\wt{\Phi}}\left(\omega;\sfrac{5}{2},\Delta\right)
        \;\; \abs{\omega}^{1/2} 
        \;\; \frac{\sqrt{6} (\sqrt{\Delta}-1)}{\pi \sqrt{1 + \sqrt{\Delta} + \Delta}} 
        \; = 1 ,\label{psdlim-alpha2-5}
        \\
        &\lim_{\Delta\rightarrow\infty} \Omega_{\wt{\Phi}}(\omega;3,\Delta)
        \;\;
        \;  2 \left[\ln{\left(1+\frac{4}{\omega^2}\right)}\right]^{-1} 
        \; = 1 \label{psdlim-alpha3}.
    \end{align}
\end{subequations}
In particular, \Eqref{psdlim-alpha2} reveals the $1/\omega$ signature of the pink noise obtained for $\alpha=2$ \cite{2005-Milotti-PhysRevE.72.056701,2023-Korzeniowska-manus-1}. 
The above equation confirms the anticipated scale invariance $1/\omega^{3-\alpha}$ of the frequency power spectral density for $1<\alpha<3$. At the boundaries of the scale-invariance regime, $\alpha=1$ and $\alpha=3$, logarithmic corrections to the frequency scaling occur. 
Similar logarithmic corrections in a closed-form power spectral density have been shown for a renewal process with power-law-distributed waiting times \cite{1993-Lowen-PhysRevE.47.992}.

Figure~\ref{fig:psd-bounded-1exp}(a) shows the compensated spectra corresponding to \Eqsref{psdlim} for a distribution width $\Delta=10^8$. 
The filled circles mark the cutoff frequencies $\omega\taumax=1$ and $\omega\taumin=1$ delimiting the theoretical frequency range of self-similar scaling for each value of $\alpha$. 
No signatures of scale invariance are exposed for the cases $\alpha=1$ and $\alpha=3$, where logarithmic corrections to the frequency scaling apply.
The $1/\omega^{3-\alpha}$ scaling is clearly manifested for the remaining Pareto exponents, $\alpha\in\{1\sfrac{1}{2},2,2\sfrac{1}{2}\}$. 
The thin dashed lines in \Figref{fig:psd-bounded-1exp}(a) mark the inverse of the $\omega$-independent factors in \Eqsref{psdlim-alpha1-5}--\eqref{psdlim-alpha2-5} indicating the constant levels which the compensated power spectral densities approach asymptotically in the limit of an infinitely broad region of scale invariance. Notably, the exposed $1/\omega^{3-\alpha}$ scaling vanishes towards the cutoff frequencies.
Figure~\ref{fig:psd-bounded-1exp}(a) demonstrates that in the case of a finite-size scale invariance, the effective frequency range of self-similar scaling is reduced by approximately two to three logarithmic decades with respect to the $\Delta$-range of the underlying distribution of pulse durations \cite{2023-Korzeniowska-manus-1}. 
This reduction is attributed to the gradual transitions between the distinct regions of a broken power law.

Figure \ref{fig:psd-bounded-1exp}(b) presents the empirical power spectral densities obtained from realizations of the process 
for bounded Pareto distributions with narrower support characterized by $\Delta=10^4$. 
The normalized sampling and duration intervals are relative to the support boundaries and equal to $\taumin/40$ and $640\taumax$, respectively.
The broken power-law transitions at $\omega\taumin=1$ and $\omega\taumax=1$ are then resolved with a one-decade margin using the Welch's method.
Each process realization comprises of more than $80\,000$ pulses, with the total number of pulses being proportional to the interval of duration and the pulse overlap parameter, here $\gamma=10$.
We recall that the analytical power spectral density expression in \Eqref{psdphifullydimless} is independent of parameter $\gamma$.
The empirical results in \Figref{fig:psd-bounded-1exp}(b) agree with the corresponding analytical predictions, including the cases where logarithmic corrections to the frequency scaling apply. 

Analogous analyses of the scale invariance of the frequency power spectral densities obtained for the upper-truncated Pareto, the lower-truncated Pareto, and the inverse Gamma distributions of pulse durations are presented in \Appref{sec:psd-other-Ptau}.
The parameter regimes for which the $1/\omega^{3-\alpha}$ scaling is manifested are summarized in Table~\ref{tab:alpha-regimes-PSD-scaling} for all the considered distribution variants.
The results are discussed in \Secref{sec:discuss-regimes}.

\begin{table}
    \centering
    \caption{Scaling signatures of the power spectral density $\Omega_{\wt{\Phi}}$ for different exponents $\alpha$ of power-law distributions of pulse durations $P_\tau$.}
    \label{tab:alpha-regimes-PSD-scaling}
    \begin{ruledtabular}
    \begin{tabular}{|c||@{}c@{}|@{}c@{}|@{}c@{}|@{}c@{}|}
        & \multicolumn{3}{c|}{ \shortstack{ $P_\tau$ Pareto \\ [0pt]} }
        & \shortstack{ $P_\tau$ inverse\\ Gamma}\\\cline{2-5} 
        &$\;\;\,\tau \in \left(0, \taumax \right]\;\;\,$
        & $\;\;\tau \in \left[\taumin, \taumax \right]\;\;$
        & $\;\;\tau \in \left[\taumin, \infty\right)\;\;$ 
        & $\;\;\tau\in(0,\infty)\;\;$
        \\ \colrule\colrule
        $0<\alpha<1$
        & \multicolumn{2}{@{}c@{}|}{\ccNoScal $\nsim 1/\omega^{3-\alpha}$} 
        & \multicolumn{2}{@{}c@{}|}{\ccNoPdf no $P_\taudim$} 
        \\ \cline{1-3} 
        $\alpha=1$
        & \ccNoPdf no $P_\taudim$
        & \cclnCorr log corr.
        & \multicolumn{2}{@{}c@{}|}{\ccNoPdf } 
        \\ \cline{1-1}\cline{3-5} 
        $1<\alpha\leq2$
        & \ccNoPdf 
        & \ccYesScal
        & \multicolumn{2}{@{}c@{}|}{\ccNosAve $\save{\taudim}=\infty$} 
        \\\cline{1-1} \cline{4-5}
        $2<\alpha<3$
        & \ccNoPdf 
        & \multicolumn{3}{@{}c@{}|}{\ccYesScal $\sim 1/\omega^{3-\alpha}$}
        \\\cline{1-1}\cline{3-5} 
        $\alpha=3$
        & \ccNoPdf 
        & \multicolumn{3}{@{}c@{}|}{ \cclnCorr log corr.}
        \\ \cline{1-1}\cline{3-5} 
        $\alpha>3$ 
        & \ccNoPdf 
        & \multicolumn{3}{@{}c@{}|}{\ccNoScal $\nsim 1/\omega^{3-\alpha}$} 
    \end{tabular}
    \end{ruledtabular}
\end{table}

\section{Discussion}\label{sec:discussion}%

The findings presented in this contribution are first related to the previous works addressing scale invariance of the frequency power spectral density for a superposition of uncorrelated pulses with distributed durations. 
The regimes of validity of the $1/\omega^{3-\alpha}$ scaling are then discussed based on the results presented in \Secsref{sec:regime-argument} and~\ref{sec:PSD}.

\subsection{Relation to previous works}

We revisit the previous pivotal works regarding spectral scaling due to distributed pulse durations and decay rates.
We highlight the elements that these works fail to address, and which we in this contribution show to be of relevance for describing the emergence of spectral scale invariance.

\subsubsection{Power spectra for distributed pulse durations and decay rates}

Distributed pulse decay rates $\rate=1/\tau$ can be equivalently considered for a process defined by \Eqref{PhiK_shotnoise}. The associated power spectral density is obtained analogously by statistical averaging, or directly form \Eqref{psdphifullydimless} using the probability of the inverse variable $P_{\rate}(\rate)$ according to \Eqref{P-inv-durdim},
\begin{align}\label{psd-inverse-ways}
    \Omega_{\wt{\Phi}}(\omega)
        = \int_0^\infty \rmd\rate\,\rate^{-2} P_{\rate}(\rate) \Varrho_\phi(\omega/\rate).
\end{align}
For $P_{\tau}(\tau)\sim \tau^{-\alpha}$ it follows that $P_\rate(\rate)\sim \rate^{-(2-\alpha)}$. Denoting ${\exprate=2-\alpha}$, the scaling relation in \Eqref{eq:beta} becomes
\begin{equation}\label{beta-rates}
    \beta=3-\alpha = 1 + \exprate.
\end{equation}
The distribution of pulse decay rates $P_\rate(\rate)$ defined on a bounded support $\rate\in[\rate_\downarrow, \rate_\uparrow]$ is for ${\exprate=0}$ reduced to a uniform distribution, which results in $1/\omega$ scaling of the power spectral density \cite{2005-Milotti-PhysRevE.72.056701}.
Obtaining a $1/\omega^\beta$ signature with $\beta\neq 1$ requires explicit power-law scalings in the distributions of pulse durations or decay rates.
Relating $P_\rate$ and $P_{\tau}$ is not generally trivial due to few known distribution pairs with explicit expressions for the forward and inverse transformations.

\subsubsection{Approach of van der Ziel}

As outlined in \Secref{sec:intro}, the approach for deriving the power spectral density of a process with distributed pulse durations from the auto-correlation function of a process where all pulses have the same duration does not account for internal long-range correlations resulting from the overlap of long-lasting, uncorrelated pulses.
Such an approach, attributed to van der Ziel \cite{1950-Ziel}, implies that the spectral scaling is proportional to $1/\omega^{2-\alpha}$ and thus that $P_\tau(\tau)\sim1/\tau$ underlies $1/f$ noise.
This result was challenged by Butz \cite{1972-Butz-JStatPhys}, and it is also incongruous with the analysis presented in this contribution demonstrating $1/\omega^{3-\alpha}$ spectral scaling of the frequency power spectral density.

Acknowledging the experimental results presented in Ref.~\onlinecite{2017-Costanzi-PhysRevLett.119.097201-1oF-superpos-Lorentzians}, speaking to the theory of van der Ziel, requires distinguishing between a superposition of interacting pulses with continuously distributed durations, and an aggregation of processes with respective constant durations.
Hooge and Bobbert argue that summation of spectra is valid only for isolated processes \cite{1994-Hooge-333808, 1997_Hooge_1997223}.
They consider an example of a generation-recombination process in an \textit{n}-type semiconductor with a range of traps whose characteristic times $\tau$ are distributed as $P_\tau(\tau)\sim 1/\tau$ in a wide range. 
The spectrum of the process is then shown to scale as $1/\omega$ only if the individual traps are isolated, while in the presence of interactions (transitions) between the traps the spectrum has Lorentzian characteristics, which agrees with the results presented here.
The authors then remark that the isolation condition is seldom considered when interpreting $1/f$ noises in electronic devices.

Unquestionably, van der Ziel and his contemporaries contributed with a large body of theoretical and experimental work to the field of flicker noise modelling \cite{1926-Schottky-PhysRev.28.74, 1925-Johnson-PhysRev.26.71, 1950-Ziel, 1979-Ziel-1979225, 1950-duPre-PhysRev.78.615, 1952-Montgomery, 1952-Petritz-4050850, 
1957-book-semiconductor-surface-physics}.
However, due to the inherent differences in the dynamics of isolated and interacting processes this theory has a limited scope of applicability.
Recognizing this is essential for selecting a representative theoretical framework.

\subsubsection{Work of Butz}

Butz derives a power spectral density integral for a process with a general pulse function and an arbitrary distribution of pulse decay rates, corresponding to \Eqref{psd-inverse-ways} \cite{1972-Butz-JStatPhys}.  
He relates pulse decay rates and durations, and demonstrates that the $1/\omega$ spectrum arises from uniformly-distributed decay rates, or equivalently from durations distributed as $1/\tau^2$.
With this he highlights the distinction with respect to the work of van der Ziel.
Butz then argues for a strict $1/\omega$ spectrum in the limit of small frequencies, or long pulse durations. 
Only asymptotic scaling relations are discussed in Ref.~\onlinecite{1972-Butz-JStatPhys} and no closed-form power spectral density expressions are offered. Therefore, there is neither any consideration of the finite-size effects discussed in this contribution.
For a power-law distribution of pulse decay rates, Butz derives a scaling relation equivalent to \Eqref{beta-rates} without considering the range of exponents for which it is valid.

\subsubsection{Work of Milotti}

Milotti considers a Poisson process as a framework for developing an exact numerical method for simulating colored noises \cite{2005-Milotti-PhysRevE.72.056701}. 
He derives a closed-form power spectral density for a superposition of pulses with a one-sided exponential shape and a bounded Pareto distribution of decay rates.
He presents the explicit expression for the case of a uniform distribution of decay rates, demonstrating the $1/\omega$ scaling.
An analogous closed-form expression was presented by van der Ziel in Ref.~\onlinecite{1979-Ziel-1979225} for $P_\tau(\tau)\sim1/\tau$, which corresponds to $P_\rate(\rate)\sim 1/\rate$ distribution of decay rates, reflecting the aforementioned discrepancies in the predicted scalings.
The closed-form power spectral densities presented by Milotti correspond respectively to the third and the second case of \Eqref{eq:psd-bounded}.

Milotti furthermore states that a scaling relation equivalent to \Eqref{beta-rates} is applicable in a range of frequencies delimited by the minimum and maximum decay rates. 
Finite-size effects associated with the cutoff frequencies are not considered. 
There is no discussion of the divergence of the lower and the upper bounds of the Pareto distribution, or the existence of the distribution and the mean duration in the relevant limits. No connection between the infinite variance of the pulse decay rate and the spectral scale invariance is made.
The exponent range in which the $\beta=1+\exprate$ scaling is valid is not considered, and no logarithmic corrections to the frequency scalings are exposed.

Milotti presents empirical power spectral densities with $1/\omega$ and $1/\omega^{1.2}$ signatures obtained for Pareto distributions spanning four logarithmic decades.
The broken power-law transitions at the minimum and maximum decay rates are not fully resolved.
The minimum distribution width necessary for a region of self-similar frequency scaling to emerge is not considered.
We recognize that the main focus of Ref.~\onlinecite{2005-Milotti-PhysRevE.72.056701} is describing the details of the numerical generator of power-law noises. The analytical results with finite-size effects presented here therefore complement and extend the works of van der Ziel, Butz and Milotti.

\subsection[Regimes of scale invariance validity]{Regimes of validity of the \texorpdfstring{\(1/\omega^{3-\alpha}\)}~ scaling}\label{sec:discuss-regimes}

The previous works considering spectral scaling for a super-position of uncorrelated pulses with distributed durations or decay rates have not addressed the conditions necessary for long-range dependence to emerge.
The regimes of validity of the $1/\omega^{3-\alpha}$ scaling are examined here.

\subsubsection{Characteristics of pulse duration distribution}

Table~\ref{tab:alpha-regimes-PSD-scaling} summarizes the scaling properties of the power spectral densities obtained for different distributions of pulse durations for different ranges of the exponent $\alpha$. 
Consistent with the expectations from Table~\ref{tab:alpha-regimes-expected-scaling},
the $1/\omega^{3-\alpha}$ scaling is manifested only if the variance of the pulse duration is infinite, which implies that the distribution allows for arbitrarily long pulse durations. This includes the bounded Pareto distribution in the limit of infinitely broad support.
Figure~\ref{fig:psd-bounded-1exp}(a) exposes an effective $1/\omega^{3-\alpha}$ scaling also for the bounded Pareto distribution with finite but wide support.
A bounded Pareto distribution with a narrow support does not result in spectral scale invariance due to finite-size effects discussed in \Secsref{sec:log-corr} and~\ref{sec:discuss-finite-size}.

A similar classification of spectral scaling characteristics was presented in Ref.~\onlinecite{2005-Lowen-book} for a filtered Poisson process with a bounded power-law pulse function.
Denoting the power-law exponent as $\alpha$,
the resulting scale invariance was shown to follow the relation $\beta=2(1-\alpha)$ for $0<\alpha<1$ given a non-divergent upper cutoff of the power-law pulse function. 
A different study addressing spectral scale invariance in fractal renewal processes with power-law distributed inter-event times demonstrated that processes with alternating values follow the $\beta=3-\alpha$ relation for $1<\alpha<3$, with logarithmic corrections to the frequency scaling at $\alpha=3$ \cite{1993-Lowen-PhysRevE.47.992}. This range of power-law exponents is the same as the one obtained in this contribution for the power-law distributed pulse durations. 
It was further shown in Ref.~\onlinecite{1993-Lowen-PhysRevE.47.992} that fractal renewal processes with Dirac delta pulses follow the $\beta=3-\alpha$ relation for $2<\alpha<3$, and the $\beta=\alpha-1$ relation for $1<\alpha<2$, with logarithmic corrections to the frequency scaling at the upper bounds of the respective ranges, $\alpha=3$ and $\alpha=2$. 
Both of the latter scaling regimes result in a range of spectral scaling exponents $\beta\in(0,1)$.

\subsubsection{Low-frequency cutoff}

The divergence of the upper bound ${\taumax\rightarrow\infty}$ entails that the region of spectral scale invariance has no cutoff at low frequencies.
The integrability of the associated power spectral density then requires $\alpha>2$, equivalent to $\beta>1$. This leads to a paradox where Parseval’s identity is seemingly violated by empirical $1/f$ noises observed in physical systems where no upper bound on the scale of events can be identified, except for the observation time itself.
The paradox has been resolved for intermittency-induced scale invariance with aging power spectra \cite{2013-Niemann-PhysRevLett.110.140603, 2015-Rodriguez-PhysRevE.92.012112, 2014-Sadegh, 2016-Leibovich-PhysRevE.94.052130}.

Unlike inter-event times, event durations are limited by finite energies fueling physical process, hence the low-frequency cutoff is always expected to occur. 
Furthermore, the paradox does not apply as soon as $\alpha>2$, for example at $\alpha=2\sfrac{1}{100}$ featured in \Figsref{fig:psd-std-invG-1exp}(a) and~\ref{fig:psd-std-invG-1exp}(c). 
The resulting spectral scaling would hardly be distinguishable from an exact $1/f$ noise in empirical power spectral densities obtained for physical measurement signals.

\subsubsection{Logarithmic corrections to the frequency scaling}\label{sec:log-corr}

Logarithmic corrections to the frequency scaling at $\alpha=1$ and $\alpha=3$ are exposed for all power spectral densities in \Secref{sec:PSD}. Spectral scale invariance is thus not manifested at the boundaries of the long-range dependence regime, $\beta=0$ and $\beta=2$.
The frequency range of self-similar scaling shortens gradually as the power-law exponent approaches ${\alpha\rightarrow1^+}$ or ${\alpha\rightarrow3^-}$. 
This is seen in \Figsref{fig:psd-std-invG-1exp}(a) and~\ref{fig:psd-std-invG-1exp}(c) where the onset of power-law scaling shifts towards lower frequencies for increasing values of $\alpha$. 
Scale invariance occurring in frequency ranges bounded on both sides vanishes completely for $\abs{\alpha-2}\gtrapprox\sfrac{6}{7}$,
where the scaling relation in \Eqref{beta-rates} effectively no longer holds \cite{2023-Korzeniowska-manus-1}.
Self-similar scaling with logarithmic corrections is then regarded as the final stage of transitioning to regular dynamics characterized by exponents $\beta\notin(0,2)$. The presence of logarithmic corrections are well known for phase transitions and critical behavior \cite{2006-Kenna-PhysRevLett.96.115701, 2010-Sandvik-PhysRevLett.104.177201, 2020-Hong-PhysRevE.101.012124}. Recently, it has also been demonstrated for the case of nonoverlapping, rectangular structures with power law distributed durations \cite{2023-Kononovicius-PhysRevE.107.034117}.

\subsubsection{Bounded scale invariance}\label{sec:discuss-finite-size}

Size limitations applying to physical systems and measurements result in truncations of the pulse duration distributions, which then manifest as broken power-law transitions in the frequency power spectral density.
Spectral characteristics outside the frequency range of scale invariance are dictated by the pulse function.
For a one-sided exponential pulse function, Lorentzian characteristics are displayed at the limiting frequencies \cite{2017-garcia-PhysPlasm}.

Figure~\ref{fig:psd-bounded-1exp}(a) demonstrates the extent to which the frequency ranges of an effective  $1/\omega^{3-\alpha}$ scaling are reduced due to the curvature associated with the broken power-law transitions.
The curvature is comparable for both the abrupt and the gradual cutoffs of the underlying distributions, as shown respectively in \Figsref{fig:psd-std-invG-1exp}(a) and~\ref{fig:psd-std-invG-1exp}(c), which are discussed in \Appref{sec:psd-other-Ptau}.
Self-similar scaling does not manifest unless at least four decades of underlying power-law scaling are present \cite{2023-Korzeniowska-manus-1}.
Logarithmic-correction effects discussed in \Secref{sec:log-corr} contribute to a further reduction of the scale invariance regions, to an extent depending on the value of $\alpha$.
The presence of persistent power-law scaling is confirmed by compensated spectra.

\subsubsection{Pulse function characteristics} \label{sec:psd-scling-other-pulse-forms}

If the exponent of pulse duration distribution does not result in a $1/\omega^{3-\alpha}$ signature, the spectral characteristics are fully determined by the pulse function.
The power spectral densities obtained for a one-sided exponential pulse function presented in \Secref{sec:PSD} demonstrate Lorentzian characteristics for $\alpha\notin[1,3]$. 
In particular, asymptotic constant scaling in the low-frequency limit, and an asymptotic $1/\omega^2$ scaling in the high-frequency limit has been analytically confirmed for: the upper-truncated Pareto distribution with $\alpha=0$; the lower-truncated Pareto and the inverse Gamma distributions with $\alpha=4$; the bounded Pareto distribution with $\alpha=0$ and $\alpha=4$ in the limits $\Delta\rightarrow1$ and $\Delta\rightarrow\infty$. 

In general, the upper bound for the range of exponents $\beta$ satisfying the $\beta=3-\alpha$ relation is determined by the pulse function and the distribution of pulse durations.
The steepness of the spectral scaling at high frequencies is determined by the highest-energy component.
A one-sided exponential pulse function contributes with $1/\omega^2$ spectral tail, 
preventing the manifestation of scaling signatures characterized by $\beta>2$.
The lower-truncated Pareto and the inverse Gamma distributions require $\alpha>2$ for a finite mean value, which further reduces the range of observable exponents to $0\leq\beta<1$. We recall that the lower bound $\beta=0$ is imposed by the condition of infinite variance of the pulse duration, $\alpha\leq3$.

Alternative pulse functions characterized by steeper spectral tails allow for manifestation of $1/\omega^{3-\alpha}$ scaling also for $\beta>2$,
if combined with pulse duration distributions whose mean value is finite for $\alpha<1$, such as the bounded Pareto distribution or the upper-truncated Pareto distribution.
Examples of relevant pulse functions include: a two-sided exponential; a Lorentzian; and a Gamma function, whose respective spectral tails scale as: $1/\omega^4$; $\exp(-2\abs{\omega})$; and $1/\omega^{2z}$, where $z$ denotes the shape parameter of the Gamma function. 
In principle, arbitrarily steep $1/\omega^{3-\alpha}$ scaling can be obtained for $\alpha<0$ and a Lorentzian pulse function.
Scale invariance characterized by exponents $\beta>2$ is, however, not associated with long-range correlations.

\section{Conclusions}\label{sec:conclusions}%

A superposition of uncorrelated pulses with durations distributed as $P_\tau(\tau)\sim1/\tau^\alpha$ displays conditional scale invariance of the frequency power spectral density of the form $1/\omega^\beta$ characterized by $\beta={3-\alpha}$.
This contribution establishes the regimes of validity of the $1/\omega^{3-\alpha}$ scaling.
We demonstrate that infinite variance of the pulse durations is necessary for the emergence of self-similar scaling associated with the long-range dependence regime $\beta\in(0,2)$.
This implies a distribution with a divergent upper bound and a finite mean value.
Closed-form power spectral density expressions derived for power-law pulse duration distributions with various truncations expose asymptotic self-similar scaling for $1<\alpha<3$, and logarithmic corrections to the frequency scaling for $\alpha=1$ and $\alpha=3$. 
Spectral characteristics for $\alpha\notin[1,3]$ are shown to be determined by the pulse function.
Analytically exposed finite-size effects associated with the bounded distribution cutoffs are shown to reduce the frequency range of self-similar scaling by two to three decades.
Consequently, spectral scale invariance does not emerge 
unless the underlying distribution of pulse durations spans at least four logarithmic decades.

We extend the work of van der Ziel, Butz and Milotti by addressing the characteristics of spectral scale invariance in the presence of size limitations for pulse durations. 
The results presented here are complementary to the work by Lowen and Teich addressing the spectral scale invariance in a Poisson process with a power-law pulse function \cite{2005-Lowen-book}, and in a renewal process with power-law-distributed inter-event times \cite{1993-Lowen-PhysRevE.47.992}.
For a superposition of localized, propagating pulses we demonstrate that power-law distributed durations $P_\tau(\tau)\sim1/\tau^2$ are equivalent to uniformly distributed velocities, motivating the emergence of power spectral density scaling as $1/f$ without an explicit underlying power-law contribution.
The presented model is highly parametrizable in terms of the admitted pulse duration distribution and the pulse function,
and it allows for generating synthetic long-range correlation processes with exactly known ranges of scale invariance and explicit power spectral density expressions. 
It is thus applicable as a framework for testing methods of spectral scaling estimation and identification of the regions of scale invariance.

\appendix

\section{Origin of power-law scalings} \label{sec:motiv} 

In this appendix it is demonstrated that a power-law distribution of pulse durations readily follows from a superposition of localized and propagating pulses with randomly distributed velocities. This motivates the investigation of power-law-distributed pulse durations as the origin of long-range correlations and power-law scaling of the frequency power spectral density.

\subsection{Motion of localized pulses} \label{sec:Pv-to-Ps}

The stochastic process given by \Eqref{PhiK_shotnoise} describing a superposition of uncorrelated pulses can be extended to include spatial advection along an axis $x$ as
\begin{equation}\label{eq:PhiK-xt}
    \Phi_{K}(x,t)=\sum_{k=1}^{K(T)} A_k\,\phi_{k}(x,t-\tarriv_{k}) ,
\end{equation}
with the pulse function at the arrival time given as
\begin{equation}
    \phi_{k}(x,0) = \psi\left(\frac{x}{\ell_{k}}\right),
\end{equation}
where $\ell_k$ is the pulse size.
The evolution of each pulse is described by an advection equation
\begin{equation}\label{eq:advc-diff}
    \frac{\partial \phi_{k}}{\partial t}+\veldim_{k} \frac{\partial \phi_{k}}{\partial x}=0,
\end{equation}
where $v_k$ is the pulse velocity. The solution of this equation is given by 
\begin{equation}
    \phi_{k}(x, t)= \psi\left(\frac{x-\veldim_{k}t}{\ell_{k}}\right).
\end{equation}
The process defined by \Eqref{eq:PhiK-xt} can accordingly be written as 
\begin{equation}
    \Phi_{K}(x, t)=\sum_{k=1}^{K(T)} A_k\, \psi\left(\frac{x-\veldim_{k}\left(t-\tarriv_{k}\right)}{\ell_{k}}\right) .
\end{equation}
At the reference position $x=0$, this reduces to the process defined by \Eqref{PhiK_shotnoise} where $\phi(\theta) = \psi(-\theta)$ and where the duration of each pulse is the transit time $\taudim=\ell/\veldim$.
A generalization of this process was used for modelling fluctuations at the boundary of magnetically confined plasmas \cite{2023-losada}.

When all pulses have the same size $\ell$, the probability density of pulse durations $P_\taudim(\taudim)$ is expressed with the probability density of pulse velocities $P_\veldim(\veldim)$ according to the the probability of the inverse variable transformation,
\begin{equation}\label{P-inv-durdim}
    P_{\taudim}(\taudim) = \frac{\ell}{\taudim^2}\, P_{\veldim}\left(\frac{\ell}{\taudim}\right).
\end{equation}
In the following, we consider both bounded and unbounded distributions of pulse velocities giving rise to power-law-distributed pulse durations, and we obtain the closed-form expressions for the resulting frequency power spectral densities.

\subsection{Uniformly distributed velocities}\label{sec:uniform-velocities}

Consider a normalized uniform distribution of pulse velocities 
\begin{equation} \label{P-unif-vel}
    \save{\veldim}\, P_{\veldim}(\veldim;\unifwidth) =
    \begin{cases}
        \displaystyle \frac{1}{2\unifwidth}  & \text{if }\veldim \in \left[\vmin, \vmax\right], \\
        0 & \text{otherwise},
    \end{cases}
\end{equation}
where $0 < \unifwidth \leq 1$ is the width of the distribution, and the minimum and the maximum velocities are given respectively as
\begin{subequations}
\begin{align}
    \vmin &=\save{\veldim}(1-\unifwidth),\\
    \vmax &=\save{\veldim}(1+\unifwidth).
\end{align}
\end{subequations}
From \Eqref{P-inv-durdim} it follows that the distribution of pulse durations has a power-law scaling,
\begin{equation} \label{eq:pareto-2}
    \save{\veldim}\, P_{\taudim}(\taudim;\unifwidth) =
    \begin{cases}
        \displaystyle \inv{2\unifwidth}\frac{\ell}{\taudim^2}  & \text{if }\taudim \in \left[\smin, \smax\right], \\
        0 & \text{otherwise},
    \end{cases}
\end{equation}
where the minimum and maximum pulse durations are respectively given by
\begin{subequations}
\begin{align}
    \smin &= \frac{\ell}{\vmax} = \frac{\ell}{\save{\veldim}(1+\unifwidth)},\\
    \smax &= \frac{\ell}{\vmin} = \frac{\ell}{\save{\veldim}(1-\unifwidth)}.
\end{align}
\end{subequations}
We define $\Delta$ to be the logarithmic width of the distribution given by \Eqref{eq:pareto-2},
\begin{equation}\label{Delta-taudim}
    \Delta=\frac{\smax}{\smin}=\frac{1+\unifwidth}{1-\unifwidth} .
\end{equation}
This definition is aligned with \Eqref{delta} for the corresponding dimensionless variable.
For any given $\Delta>1$, the width parameter is given by ${\unifwidth=(\Delta-1)/(\Delta+1)}$, and the average pulse duration then becomes
\begin{equation}\label{eq:meantau-pareto-2}
    \save{\taudim}=\int_{\smin}^{\smax} \rmd \taudim\, \taudim\, P_{\taudim}(\taudim;\unifwidth)
    = \frac{\ell}{2\save{\veldim}}\frac{\Delta+1}{\Delta-1}\, \ln{\Delta} .
\end{equation}
This mean value diverges logarithmically in the limit $\Delta\rightarrow\infty$. 

In the limit of the widest possible pulse velocity distribution, $\unifwidth\rightarrow 1$, the minimum velocity vanishes, $\vmin\rightarrow0$. Arbitrarily slow pulses lead to excessively long pulse durations, and consequently to the divergence of the average pulse duration $\save{\taudim}$ and the maximum duration $\smax$, while the minimum duration $\smin$ remains finite,
\begin{equation}\label{unifdistlimits}
    \lim_{\unifwidth \rightarrow 1} \smin = \frac{\ell}{2\save{\veldim}} < \infty ,
    \quad
    \lim_{\unifwidth \rightarrow 1} \smax = \infty,
    \quad
    \lim _{\unifwidth \rightarrow 1} \save{\taudim} = \infty .
\end{equation}
As will be discussed presently, this motivates consideration of a lower-truncated Pareto distribution for pulse durations.

\subsection{Pareto distributed durations}\label{sec:motiv-pareto}

Consider now a generalized bounded Pareto distribution of pulse durations with power-law exponent $\alpha$,
\begin{equation} \label{gpareto-b}
    P_\taudim(\taudim;\alpha,\smin,\smax) =
    \begin{cases}
        \displaystyle \frac{(\alpha-1)\smin^{\alpha-1}}{1 - \Delta^{1-\alpha}}\taudim^{-\alpha}  & \text{if }\taudim \in \left[\smin, \smax\right], \\
        0 & \text{otherwise},
    \end{cases}
\end{equation}
whose mean value is given by
\begin{equation}\label{eq:avetaudim}
    \save{\taudim} = \frac{(\alpha-1)}{(\alpha-2)}
    \frac{(1-\Delta^{2-\alpha})}{(1-\Delta^{1-\alpha})}\, \smin.
\end{equation}
When $\alpha=2$, the Pareto distribution in \Eqref{gpareto-b} simplifies to the distribution given by \Eqref{eq:pareto-2}. The corresponding mean value in the limit $\alpha\rightarrow2$ becomes
\begin{equation}\label{eq:lim2-avetaudim}
    \lim_{\alpha\rightarrow2} \save{\taudim} = \frac{\Delta\,\ln\Delta}{\Delta-1} \smin
    = \frac{\ell}{2\save{\veldim}}\frac{\Delta+1}{\Delta-1}\ln \Delta ,
\end{equation}
which coincides with the mean value in \Eqref{eq:meantau-pareto-2}.
In \Secref{sec:analytic-PSD} it is demonstrated that for $\alpha=2$ the frequency power spectral density scales as the inverse of the frequency, giving rise to the $1/f$-noise pattern.
This motivates consideration of the generalized bounded Pareto distribution of pulse durations scaling as $P_\taudim(\taudim) \sim 1/{\taudim^\alpha}$ in the range $\taudim\in[\smin,\smax]$, with a continuous range of exponents $\alpha$ resulting in a range of scaling signatures of the corresponding frequency power spectral density. The details of such a distribution are addressed in \Secref{sec:Pareto-bounded}.

\begin{figure}
    \includegraphics{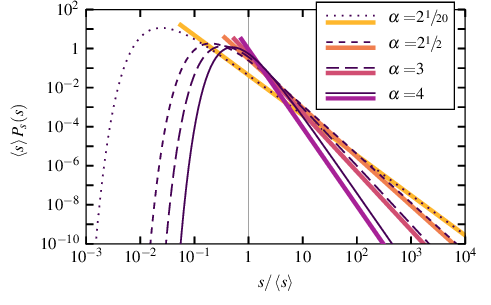}
    \caption{Probability density functions of the lower-truncated Pareto distribution in \Eqref{gpareto-s} 
    (thick solid lines) and the inverse Gamma distribution in \Eqref{eq:inv-gamma}
    (thin featured lines).
    The pulse durations are normalized with the mean value $\save{\taudim}$. 
    }
    \label{fig:cmp_stdPareto-invGamma}
\end{figure}

The limits of the pulse durations in \Eqref{unifdistlimits} for a uniform velocity distribution motivates consideration of the lower-truncated Pareto distribution of pulse durations in the range $\taudim\in[\smin, \infty)$, given by
\begin{equation} \label{gpareto-s}
    P_\taudim(\taudim;\alpha,\smin) =
    \begin{cases}
        \displaystyle (\alpha-1)\, \smin^{\alpha-1}\,\taudim^{-\alpha}  & \text{if }\taudim \in \left[\smin, \infty\right), \\
        0 & \text{otherwise} ,
    \end{cases}
\end{equation}
which is well-defined for $\alpha>1$. 
The mean value is finite for $\alpha>2$ and then given by
\begin{equation}\label{std-par-mean}
    \save{\taudim}  = \frac{\alpha - 1}{\alpha - 2} \smin.
\end{equation}
The lower-truncated Pareto distribution for pulse durations is presented in \Figref{fig:cmp_stdPareto-invGamma} for various values of the scaling exponent $\alpha$. Further details on the lower-truncated Pareto distribution in \Eqref{gpareto-s} and the resulting frequency power spectral density are presented in \Appref{sec:Pareto-standard}.

\begin{figure*}
    \includegraphics[width=0.5\textwidth]{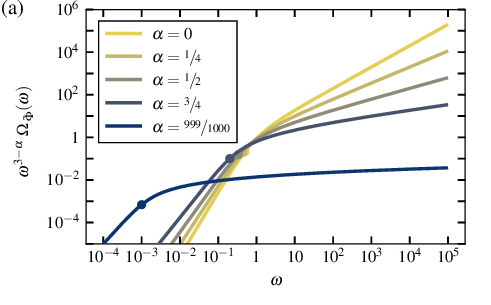}%
    \includegraphics[width=0.5\textwidth]{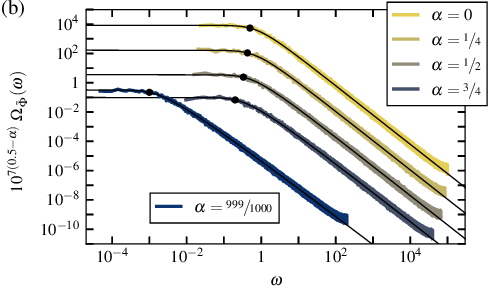}
    \caption{
        Power spectral densities for a super-position of one-sided exponential pulses with upper-truncated-Pareto distributed durations. 
        Filled circles mark the cutoff frequencies $\omega\taumax=1$.
        The presented values of $\alpha$ span the entire range where the distribution is well-defined.
        (a) Compensated analytical power spectral densities $\omega^{3-\alpha}\,\Omega_{\wt{\Phi}}(\omega; \alpha)$ with $\Omega_{\wt{\Phi}}$ given by \Eqref{eq:psd-upper-trunc}.
        (b) Empirical power spectral densities for realizations of the process.
        The respective analytical power spectral densities are plotted with solid black curves. 
        Vertical shifting by $\alpha$-dependent factors is applied to avoid overlapping.
        }
    \label{fig:psd-upper-truncated-1exp}
\end{figure*}

\subsection{Gamma distributed velocities}\label{sec:gamma-velocities}

Consider now a Gamma distribution of pulse velocities with shape parameter $\alpha-1$ and scale parameter $\save{\veldim}/(\alpha-1)$, which for $\veldim>0$ is given by
\begin{equation} \label{ggamma}
    \save{\veldim}\,P_{\veldim}(\veldim;\alpha) =
        \frac{(\alpha-1)^{\alpha}}{\Gamma(\alpha)} \left(\frac{\veldim}{\save{\veldim}}\right)^{\alpha-2} \exp\left(\frac{(1-\alpha)\veldim}{\save{\veldim}}\right) .
\end{equation}
For $\alpha>2$, this distribution is unimodal and has an exponential tail. 
For large values of $\alpha$ it resembles a narrow normal distribution, corresponding to the case where all pulses have the same velocity.

According to \Eqref{P-inv-durdim}, the distribution of pulse durations is in this case given by the inverse Gamma distribution,
\begin{equation}\label{eq:inv-gamma} 
     P_\taudim(\taudim;\alpha) = \frac{\save{\veldim}}{\ell} \frac{(\alpha-1)^{\alpha}}{\Gamma(\alpha)} \left(\frac{\ell}{\save{\veldim}\taudim}\right)^{\alpha} \exp\left(\frac{(1-\alpha)\ell}{\save{\veldim}\taudim}\right) ,
\end{equation}
with an average pulse duration for $\alpha>2$ given by
\begin{equation}\label{mean-dim-inv-gamma}
    \save{\taudim}=\frac{\ell}{\save{\veldim}}\frac{\alpha-1}{\alpha-2} .
\end{equation}
This is equivalent to the mean value in \Eqref{std-par-mean} for the lower-truncated Pareto distribution, up to the $\alpha$-independent prefactor. 

At relatively high values of $\taudim$, the inverse Gamma distribution in \Eqref{eq:inv-gamma} follows the same power-law scaling $P_{\taudim}(\taudim)\sim\taudim^{-\alpha}$ as the lower-truncated Pareto distribution in \Eqref{gpareto-s}.
Figure~\ref{fig:cmp_stdPareto-invGamma} reveals the parallel tails of the corresponding distributions
for several values of $\alpha$.
The difference with respect to the lower-truncated Pareto distribution is that the inverse Gamma distribution is defined on the entire positive range $\taudim\in(0,\infty)$, without a cutoff at $\smin$. Instead, it has a relatively sharp but smooth drop at low values of $\taudim$, as demonstrated in \Figref{fig:cmp_stdPareto-invGamma}. A gradual cutoff may be considered more representative of physical finite-size power-law scaling than the abrupt cutoff of the lower-truncated Pareto distribution given by \Eqref{gpareto-s}.
Further details on the inverse Gamma distribution for pulse durations and the resulting frequency power spectral density are presented in \Appref{sec:inverse-Gamma-distr}.

\section{Power spectral densities for alternative power-law distributions of pulse durations}\label{sec:psd-other-Ptau}

This appendix is an extension of \Secsref{sec:distributed-Ptau} and~\ref{sec:PSD}.
It presents the definitions of the remaining variants of power-law distributions, complementary to \Secref{sec:Pareto-bounded}, followed by the derivation and analysis of the associated power spectral densities, complementary to \Secref{sec:analytic-PSD}.

\subsection{Upper-truncated Pareto distribution}\label{sec:Pareto-upper-truncated}

The Pareto distribution defined by \Eqref{gpareto} with an upper-truncated support $\tau \in \left(0, \taumax \right]$ is well-defined for $\alpha < 1$. Applying conditions \eqref{normalized_Ptau} and~\eqref{normalized_taumean} yields the following parameters of the distribution,
\begin{align}
    \eta (\alpha, \taumax) &= -(\alpha-1)\, \taumax^{\alpha-1}, \label{eta-upper-tr-Pareto}\\
    \taumax(\alpha) &= \frac{\alpha-2}{\alpha-1}.\label{upper-tr-Pareto-last}
\end{align}
The pulse duration variance is finite for $\alpha<1$ and then given by
\begin{equation}
    \save{\tau^2}(\alpha) = \frac{\alpha-2}{\alpha-3} \taumax. 
\end{equation}
The upper-truncated Pareto distribution was not motivated in \Appref{sec:motiv}.

For the upper-truncated Pareto distribution characterized by \Eqsref{gpareto} and~\eqref{eta-upper-tr-Pareto}--\eqref{upper-tr-Pareto-last}, the power spectral density given by \Eqref{psdphifullydimless} for $0<\alpha<1$ becomes 
\begin{equation}\label{eq:psd-upper-trunc}
    \Omega_{\wt{\Phi}}(\omega;\alpha) =\, \frac{2(\alpha-2)}{(\alpha-3)}\, _2F_1\left(1,\frac{3-\alpha}{2},\frac{5-\alpha}{2};-\taumax^2\omega^2 \right) \,\taumax ,
\end{equation}
where $\taumax$ is given by \Eqref{upper-tr-Pareto-last} and $_2F_1$ is a hypergeometric function defined by Gauss series \cite{hyp2f1}.
As discussed in \Secref{sec:regime-argument}, spectral scale invariance is expected only in the limit $\alpha\rightarrow1^-$, in which case the asymptotic Lorentzian tail $\sim1/\omega^2$ results from the discontinuity in the pulse function rather than from self-similarity of the process. 
The lack of self-similar scaling is confirmed in \Figref{fig:psd-upper-truncated-1exp}(a), which shows that the power spectral density given by \Eqref{eq:psd-upper-trunc} compensated by a factor $\omega^{3-\alpha}$ does not maintain a constant value in any frequency range. Hence, $\Omega_{\wt{\Phi}}\nsim1/\omega^{3-\alpha}$ for $0<\alpha<1$. 
Figure~\ref{fig:psd-upper-truncated-1exp}(b) presents the empirical power spectral densities obtained for realizations of the process for $\gamma = 10$. The normalized sampling and duration intervals are given by $\taumax/(4\times10^5)$ and $640 \taumax$, respectively. 
The empirical results demonstrate perfect agreement with the corresponding analytical predictions, and show that a Lorentzian shape is obtained irrespectively of the value of $\alpha$.

\begin{figure*}
    \includegraphics[width=0.5\textwidth]{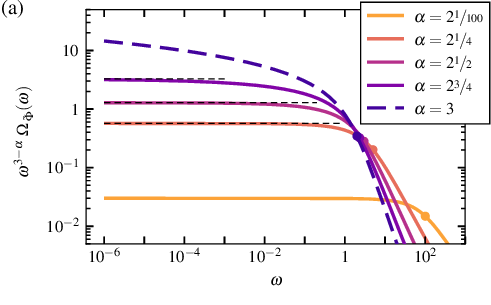}%
    \includegraphics[width=0.5\textwidth]{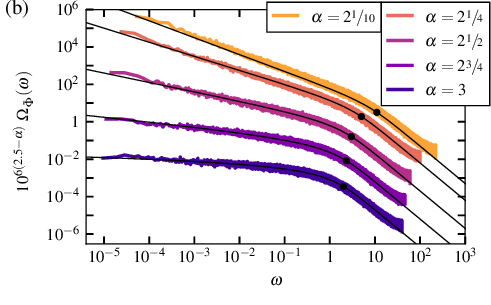}
    \\
    \includegraphics[width=0.5\textwidth]{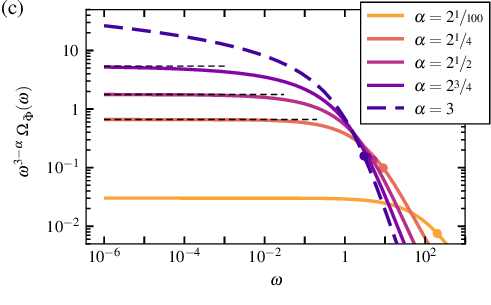}%
    \includegraphics[width=0.5\textwidth]{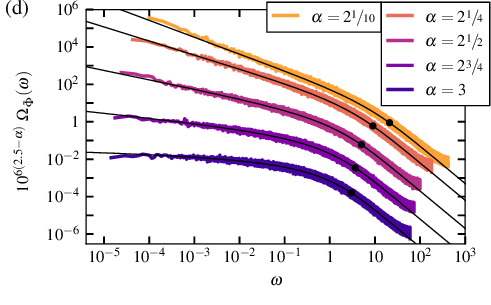}
    \caption{
        Power spectral densities for a super-position of one-sided exponential pulses with distributed durations: (top row) lower-truncated Pareto distribution with the cutoff frequency $\omega\taumin=1$; (bottom row) inverse Gamma distribution with the cutoff frequency $\omega\tau_m=1$. 
        Explanation of symbols as in \Figref{fig:psd-bounded-1exp}.
        Left column: Compensated analytical power spectral densities $\omega^{3-\alpha}\,\Omega_{\wt{\Phi}}(\omega; \alpha)$ with $\Omega_{\wt{\Phi}}$ given by: (a) \Eqref{eq:psd-standard}; (c) \Eqref{eq:psd-inv-gamma}, with $\omega$-independent factors in: (a) \Eqsref{eq:compens-psd-standard}; 
        (c) \Eqsref{eq:compens-psd-inv-G-225}--\eqref{eq:compens-psd-inv-G-275}. 
        Right column: Empirical power spectral densities for realizations of the process.
        The respective analytical power spectral densities are plotted with solid black curves. 
        Vertical shifting by $\alpha$-dependent factors is applied to avoid overlapping.
        }
    \label{fig:psd-std-invG-1exp}
\end{figure*}

\subsection{Lower-truncated Pareto distribution}\label{sec:Pareto-standard}

The Pareto distribution defined by \Eqref{gpareto} with a lower-truncated support $\tau \in \left[\taumin, \infty\right)$ is well-defined for $\alpha > 2$. This corresponds to the standard Pareto distribution. Applying conditions \eqref{normalized_Ptau} and~\eqref{normalized_taumean} yields the following parameters of the distribution,
\begin{align}
    \eta (\alpha, \taumin) &= (\alpha-1)\, \taumin^{\alpha-1},\label{eta-lower-tr-Pareto}\\
    \taumin(\alpha) & = \frac{\alpha-2}{\alpha-1}.\label{lower-tr-Pareto-last}
\end{align}
The pulse duration variance is finite for $\alpha>3$ and then given by 
\begin{equation}
    \save{\tau^2}(\alpha) =  \frac{\alpha-2}{\alpha-3} \taumin .
\end{equation}
This definition of the lower-truncated Pareto distribution is aligned with \Eqref{gpareto-s} motivated in \Appref{sec:uniform-velocities}.

For the lower-truncated Pareto distribution characterized by \Eqsref{gpareto} and~\eqref{eta-lower-tr-Pareto}--\eqref{lower-tr-Pareto-last}, the power spectral density given by \Eqref{psdphifullydimless} for $\alpha>2$ becomes 
\begin{align}\label{eq:psd-standard}
    \Omega_{\wt{\Phi}}(\omega;\alpha) =
    \frac{2}{\omega^2}\, _2F_1\left(1,\frac{\alpha-1}{2},\frac{\alpha+1}{2};\frac{-1}{\taumin^2\omega^2} \right) ,
\end{align}
where $\taumin$ is given by \Eqref{lower-tr-Pareto-last}.
Spectral scale invariance is expected for $2<\alpha\leq3$. 
Compensating \Eqref{eq:psd-standard} by $\omega^{3-\alpha}$ exposes self-similar scaling in the low-frequency limit,
\begin{subequations}\label{eq:compens-psd-standard}
    \begin{align}
        &\lim_{\omega\rightarrow0}\, \Omega_{\wt{\Phi}}\left(\omega;2\sfrac{1}{4}\right) 
        \; \abs{\omega}^{3/4} 
        \; \frac{2}{\pi} (30 + 20 \sqrt{2})^{1/4}
        \; = 1,
        \\
        &\lim_{\omega\rightarrow0}\, \Omega_{\wt{\Phi}}\left(\omega;2\sfrac{1}{2}\right) 
        \; \abs{\omega}^{1/2} 
        \; \frac{\sqrt{6}}{\pi}
        \; = 1,
        \\
        &\lim_{\omega\rightarrow0}\, \Omega_{\wt{\Phi}}\left(\omega;2\sfrac{3}{4}\right)
        \; \abs{\omega}^{1/4} 
        \; \frac{2}{3\pi}\left(\frac{7}{3}\right)^{3/4} 
        \left( {1+\frac{1}{\sqrt{2}}}\right)^{-1/2} 
        \; = 1.
\end{align}    
\end{subequations}
The inverse of the $\omega$-independent compensating factors in \Eqsref{eq:compens-psd-standard} indicate the levels to which the compensated power spectral density tends asymptotically in the low-frequency limit. These levels are marked in \Figref{fig:psd-std-invG-1exp}(a) with horizontal dashed black lines. The frequency ranges where the compensated-spectra curves overlap with the constant-level lines indicate regions of $1/\omega^{3-\alpha}$ scaling. 
Figure~\ref{fig:psd-std-invG-1exp}(a) demonstrates also that the compensated spectra for $\alpha\rightarrow2^+$ manifest a clear signature of scale invariance
\footnote{
For $1<\alpha\leq2$ the mean value for the dimensional lower-truncated Pareto distribution is infinite, $\save{\taudim}=\infty$, preventing the normalization of the power spectral density into the form given by \Eqref{psdphifullydimless}.
Nevertheless, the dimensional power spectral density integral in \Eqref{psdPhiTilde} can be solved for a generic lower-truncated Pareto distribution defined on a support $\taudim\in[\smin, \infty)$ as $P_\taudim(\taudim)=\eta\, \taudim^{-2}$, 
\begin{equation}\label{eq:not-normalized-psd-standard}
    \int_{\smin}^\infty \rmd\taudim\,\taudim^2 P_{\taudim}(\taudim) \Varrho_\phi(\taudim\omegadim)
            = \eta \left(\frac{\pi}{\abs{\omegadim}} - \frac{2\arctan\left(\omegadim\smin\right)}{\omegadim}\right) ,
\end{equation}
which in the limit $\smin\rightarrow0$ integrates to $\eta\pi/\abs{\omegadim}$,
revealing the expected $1/\omegadim$ scaling, in line with the $\beta=3-\alpha$ relation for $\alpha=2$.
}.

For $\alpha=3$ the power spectral density in \Eqref{eq:psd-standard} simplifies to
\begin{equation}\label{eq:psd-a3-std}
    \Omega_{\wt{\Phi}}(\omega;3)
                = \frac{1}{2}\ln\left(1+\frac{4}{\omega^2}\right),
\end{equation}
revealing logarithmic corrections to the frequency scaling. The compensated spectra in \Figref{fig:psd-std-invG-1exp}(a) confirm the lack of self-similar scaling for $\alpha=3$. 
The corresponding empirical power spectral densities presented in \Figref{fig:psd-std-invG-1exp}(b) are aligned with the analytical expressions.
The underlying process realizations are obtained for $\gamma = 10$. The normalized sampling and duration intervals are given by $\taumin/40$ and $64\times10^5\,\taumin$, respectively.

\subsection{Inverse Gamma distribution}\label{sec:inverse-Gamma-distr}

The inverse Gamma distribution for the normalized pulse durations follows from \Eqref{eq:inv-gamma}. It is well-defined for $\alpha>2$ and then given by
\begin{equation}\label{eq:inv-gamma-distr-norm-alpha}
    P_{\tau}(\tau;\alpha) = \frac{(\alpha-2)^{\alpha-1}}{\Gamma(\alpha-1)} \tau^{-\alpha} \exp\left(\frac{2-\alpha}{\tau}\right) .
\end{equation}
This distribution has a mode at 
\begin{equation}\label{eq:inv-Gamma-mode}
    \tau_\text{m} = \frac{\alpha-2}{\alpha},
\end{equation}
which marks the onset of the smooth exponential decrease to zero value in the limit $\tau\rightarrow0$, as shown in \Figref{fig:cmp_stdPareto-invGamma}.

The parametrization $P_\tau(\tau;\alpha)$ of the inverse Gamma distribution in \Eqref{eq:inv-gamma-distr-norm-alpha} aligns it to the power-law notation $\tau^{-\alpha}$ used for the Pareto distributions introduced in \Secref{sec:variants-Pareto}. 
The inverse Gamma distribution has the same support span $\tau\in(0,\infty)$ as the ill-defined unbounded Pareto distribution. However, due to the smooth cutoff at low values of $\tau$ resulting from the exponential term, it is both well-defined according to \Eqref{normalized_Ptau}, and it has a finite, non-diverging mean satisfying the normalization condition \eqref{normalized_taumean}. 

As highlighted in \Appref{sec:gamma-velocities}, the inverse Gamma distribution is qualitatively comparable to the lower-truncated Pareto distribution: it has the same scaling for the tail, effectively a cutoff at low pulse durations, and a finite pulse duration variance for $\alpha>3$, which for the inverse Gamma distribution is given by
\begin{equation}
    \save{\tau^2}(\alpha) = \frac{\alpha-2}{\alpha-3} .
\end{equation}

The power spectral density in \Eqref{psdphifullydimless} is for the inverse Gamma distribution in \Eqref{eq:inv-gamma-distr-norm-alpha} with $\alpha>2$ given by
\begin{multline} \label{eq:psd-inv-gamma}
    \Omega_{\wt{\Phi}}(\omega;\alpha) = \\
    \begin{cases}
        \sin(\abs{\omega})\left[\pi - 2\,\text{Si}(\abs{\omega})\right] -2\cos(\omega)\, \text{Ci}(\abs{\omega}) & \text{if } \alpha = 3, \\
        \frac{2}{\Gamma(\alpha-1)} 
        \left[  \abs{\omega}^{\alpha-3} 
                \cos\left(\frac{\pi(\alpha-1)}{2}+(\alpha-2)\abs{\omega}\right)
                \right.\\ 
                \qquad\qquad\times 
                \pi(\alpha-2)^{\alpha-1} 
                \csc(\pi (\alpha-1)) \\
            \quad\qquad + 
                _pF_q\left(1,\left\{\frac{4-\alpha}{2},\frac{5-\alpha}{2}\right\},\frac{(\alpha-2)^2}{-4}\omega^2\right) \\
                \left. \qquad\qquad\times(\alpha-2)^2\, \Gamma(\alpha-3)\right] & \text{otherwise},
    \end{cases} 
\end{multline}
where $\text{Si}(z)=\int_0^z \rmd t\,\sin(t)/t$ and $\text{Ci}(z)=-\int_z^\infty \rmd t\,\cos(t)/t$ are sine and cosine integrals, respectively, and $_pF_q$ is a generalized hypergeometric function defined by generalized hypergeometric series \cite{hyppfq}. Compensating \Eqref{eq:psd-inv-gamma} by $\omega^{3-\alpha}$ reveals the expected scaling in the low-frequency limit,
\begin{subequations}
    \begin{align}
        &\lim_{\omega\rightarrow0} \Omega_{\wt{\Phi}}\left(\omega;2\sfrac{1}{4}\right) 
        \; \abs{\omega}^{3/4} 
        \;\; \frac{2\;\Gamma\left(\frac{5}{4}\right)}{\pi\,\sin\left(\frac{\pi}{8}\right)}
        \; = 1 ,\label{eq:compens-psd-inv-G-225}\\
        &\lim_{\omega\rightarrow0} \Omega_{\wt{\Phi}}\left(\omega;2\sfrac{1}{2}\right) 
        \; \abs{\omega}^{1/2} 
        \;\; \frac{1}{\sqrt{\pi}}
        \;= 1 ,\label{eq:compens-psd-inv-G-250}\\
        &\lim_{\omega\rightarrow0} \Omega_{\wt{\Phi}}\left(\omega;2\sfrac{3}{4}\right) 
        \; \abs{\omega}^{1/4} 
        \;\; \frac{4\sqrt{2}\;\Gamma\left(\frac{7}{4}\right)}{\pi\, 3^{7/4}\sqrt{1+\frac{1}{\sqrt{2}}}}
        \;= 1 ,\label{eq:compens-psd-inv-G-275}\\
        &\lim_{\omega\rightarrow0} \Omega_{\wt{\Phi}}(\omega;3)
        \;
        \;\left[\ln{\left(1+\frac{4}{\omega^2}\right)}\right]^{-1} 
        \;= 1.\label{eq:compens-psd-inv-G-3}
    \end{align}
\end{subequations}
For $\alpha=3$ the same logarithmic correction to the frequency scaling is present as for the power spectral densities in \Eqsref{eq:psd-a3-std} and~\eqref{psdlim-alpha3} derived for the lower-truncated Pareto and the bounded Pareto distributions, respectively. 
For $\alpha=2$ the inverse Gamma distribution vanishes and so does the power spectral density given by \Eqref{eq:psd-inv-gamma}. 

Figure~\ref{fig:psd-std-invG-1exp}(c) confirms the presence of the $1/\omega^{3-\alpha}$ scaling for $2<\alpha<3$, and the lack thereof for $\alpha=3$. 
Figure~\ref{fig:psd-std-invG-1exp}(d) shows the empirical power spectral densities obtained for realizations of the process for $\gamma = 10$. The normalized sampling and duration intervals are given by $\tau_m/40$ and $64\times10^5\, \tau_m$, respectively.
Expectedly, the empirical results match the analytical expressions.
Figure \ref{fig:psd-std-invG-1exp} demonstrates also that the scaling signatures of the frequency power spectral density obtained for the corresponding values of $\alpha$ are remarkably similar between the inverse Gamma distribution with a smooth cutoff at $\tau_m$, and the lower-truncated Pareto distribution with an abrupt cutoff at $\taumin$.

\begin{acknowledgments}
    This work was supported by the UiT Aurora Centre Program, UiT The Arctic University of Norway (2020). Discussions with Martin Rypdal and Audun Theodorsen are gratefully appreciated.
\end{acknowledgments}

%

\end{document}